\documentclass[prd,amssymb,aps,twocolumn,epsfig]{revtex4}
\usepackage{graphicx}% Include figure files
\usepackage{dcolumn}   % needed for some tables
\usepackage{amsmath,amssymb}
\usepackage{bm}
\usepackage[colorlinks,
              linkcolor=blue,
              citecolor=blue,
              urlcolor=blue,
              anchorcolor=blue]{hyperref}

\def\kc#1{\left(#1\right)}

\def\be{\begin{equation}}       \def\ee{\end{equation}}
\def\bea{\begin{eqnarray}}      \def\eea{\end{eqnarray}}
\def\ba{\begin{array} }
\def\ea{\end{array} }
\def\bc{\begin{center}}
\def\ec{\end{center}}
\def\bnum{\begin{enumerate} }
\def\enum{\end{enumerate}}

\def\=>{\Rightarrow}
\def\>{\rightarrow}

\date{\today}

\begin{document}

\title{Model-independent probe of anomalous heavy neutral Higgs bosons at the LHC }

\author{Yu-Ping Kuang$^{1,2}$\footnote{ypkuang@mail.tsinghua.edu.cn},
Hong-Yu Ren$^{1}$\footnote{renhy10@mails.tsinghua.edu.cn}, and
Ling-Hao Xia$^{1}$\footnote{xlh10@mails.tsinghua.edu.cn}}

\affiliation{$^1$ Department of Physics, Tsinghua University,
Beijing, 100084, China}

\affiliation{$^2$ Center for High Energy Physics, Tsinghua
University, Beijing, 100084, China}

\begin{abstract}

We first formulate, in the framework of effective Lagrangian, the
general form of the effective interactions of the lightest Higgs
boson $h$ and a heavier neutral Higgs boson $H$ in a multi-Higgs
system taking account of Higgs mixing effect. We regard $h$ as the
discovered Higgs boson which has been shown to be consistent with
the standard model (SM) Higgs boson. The obtained effective
interactions contain extra parameters reflecting the Higgs mixing
effect. Next, We study the constraints on the anomalous coupling
constants of $H$ from both the requirement of the unitarity of the
$S$-matrix and the exclusion bounds on the SM Higgs boson obtained
from the experimental data at the 7--8 TeV LHC. From this we
obtain the available range of the anomalous coupling constants of
$H$, with which $H$ is not excluded by the yet known theoretical
and experimental constraints. We then study the signatures of $H$
at the 14 TeV LHC. In this paper, we suggest taking weak-boson
scattering and $pp\to VH^\ast\to VVV$ as sensitive processes for
probing $H$ model independently at the 14 TeV LHC. We take several
examples with the anomalous $HVV$ coupling constants in the
available ranges to do the numerical study. a full tree-level
calculation at the hadron level is given with signals and
backgrounds carefully calculated. We impose a series of proper
kinematic cuts to effectively suppress the backgrounds. It is
shown that, in both the $VV$ scattering and the $pp\to VH^\ast\to
VVV$ processes, $H$ boson can be discovered from the invariant
mass distributions of the final state particles with reasonable
integrated luminosity. Especially, in the $pp\to VH^\ast\to VVV$
process, the invariant mass distribution
%$M(J_1,J_2)$
of the final state jets can show a clear resonance peak of $H$.
%at $M(J_1,J_2)=M^{}_H$.
Finally, we propose several physical observables from which the
values of the anomalous coupling constants $f^{}_W$ and
$f^{}_{WW}$ can be measured
experimentally.\\\\

\null\noindent{PACS numbers: 14.80.Ec, 12.60.Fr, 12.15.-y}
\end{abstract}

\null\noindent{\null\hspace{5cm}TUHEP-TH-14180\&1}

\maketitle

\section{Introduction}

The discovery of the 125--126 GeV Higgs boson \cite{ATLAS_CMS12}
at the LHC in 2012 is a milestone in our understanding of the
electroweak (EW) theory. So far, the measured gauge and Yukawa
couplings of this 125--126 GeV Higgs boson are consistent with the
standard model (SM) couplings \cite{CMS_JHEP13&ATLAS1305}. Since
the precision of the present measurements at the LHC is still
rather mild due to the large hadronic backgrounds, a new high
energy electron-positron collider is expected for higher precision
measurements of the Higgs properties \cite{EP}. However, even if
the measured precise values of the 125--126 GeV Higgs boson
couplings are very close to the SM values, it does not imply that
the SM is a final theory of fundamental interactions since the SM
suffers from various shortcomings, such as the well-known
theoretical problems of {\it triviality} \cite{triviality} and
{\it unnaturalness} \cite{unnaturalness}; the facts that it does
not include the dark matter; it can neither predict the mass of
the Higgs boson nor predict the masses of all the fermions, etc.
Searching for new physics beyond the SM is the most important goal
of future particle physics studies.

Most new physics models contain more than one Higgs bosons. In
many well-known new physics models (such as the two-Higgs-doublet
models (2HDM), the minimal supersymmetric extension of the SM
(MSSM), the left-right symmetric models, etc), the lightest Higgs
boson may behave rather like a SM Higgs boson, and the masses of
other heavy Higgs bosons are usually in the few hundred GeV to TeV
range. So it is quite possible that the discovered 125--126 GeV
Higgs is the lightest Higgs boson in certain new physics models.
Since the few hundred GeV to TeV range is within the searching
ability of the LHC, searching for non-standard (NS) heavy neutral
Higgs bosons at the 14 TeV LHC is thus a feasible way of finding
out the correct new physics model beyond the SM.

There are a lot of proposed new physics models in the literatures
in which the Higgs bosons can be either elementary or composite,
and we actually do not know whether the correct new physics model
reflects the nature is just one of these proposed models or not.
Therefore just searching for heavy Higgs bosons model by model at
the LHC is not an effective way. For example, there have been
experimental searches for the heavy Higgs bosons in the MSSM and
the 2HDM with negative results
\cite{CMS_1201.4893}\cite{ATLAS_1305.3315}\cite{ATLAS_PRD89}. A
more effective way is to perform a general search for the heavy
neutral Higgs boson model independently.

In the following, we shall treat the discovered 125-126 GeV Higgs
boson as a SM-like Higgs with negligible anomalous couplings
\cite{Eboli}. For a neutral heavier Higgs boson with not so small
gauge interactions (there may be gauge-phobic heavy neutral Higgs
bosons which are not considered in the present study), we shall
give a general model-independent formulation of the gauge and
Yukawa couplings of the NS heavy neutral Higgs boson in a
multi-Higgs system taking account of the Higgs mixing effect based
on the effective Lagrangian consideration, which contains several
unknown coupling constants. We then study the constraints on the
unknown coupling constants both theoretically and experimentally.
We shall first study the theoretical upper bounds on these unknown
coupling constants from the requirement of the unitarity of the
$S$-matrix. Then we shall consider the $95\%$ CL experimental
exclusion limits on the SM Higgs boson obtained from the CMS
(ATLAS) data at the 7--8 TeV LHC
\cite{CMS_HIG_13_002}-\cite{CMS13008}. The condition for the NS
heavy neutral Higgs bosons to avoid being excluded is that they
should have large enough anomalous couplings to sufficiently
reduce their production rates. %This will give certain lower bounds
%on the unknown coupling constants.
These bounds provide certain knowledge on the possible range of
these unknown coupling constants, which can be a starting point of
our study of the model-independent detection of the NS heavy
neutral Higgs boson $H$ at the LHC.

In this paper, we consider a general multi-Higgs system with Higgs
mixing caused by the general multi-Higgs interactions. In the mass
eigenstate, we pay special attention to the lightest Higgs boson
$h$ and the heavier Higgs boson (heavier than $h$ but lighter than
other heavy Higgs bosons) $H$. We regard $h$ as the discovered
125--126 GeV Higgs boson which has been shown to be consistent
with the SM Higgs boson. We then formulate the effective
interactions related to $h$ and $H$ up to the dim-6 operators.
Since $h$ is consistent with the SM Higgs boson, we neglect its
dim-6 interactions. The obtained effective interactions are
different from the conventional one constructed for a single-Higgs
system \cite{Hagiwara} by containing extra new parameters
reflecting Higgs mixing effect.

Next, we study the existing theoretical and experimental
constraints on the parameters in the effective interactions.
Theoretically, we require the present theory does not violate the
unitarity of the $S$-matrix. Experimentally, we require the heavy
Higgs boson $H$ is not excluded by the CMS (ATLAS) exclusion bound
on the SM Higgs boson \cite{CMS_HIG_13_002}. These constraints
determine an {\it available} region for the anomalous coupling
constants with which the heavy Higgs boson $H$ is not excluded by
the present theoretical and experimental requirements. This
provides the staring point of studying the model-independent probe
of the heavy Higgs boson $H$ at the 14 TeV LHC.

In this paper, we suggest taking weak-boson scattering and $pp\to
VH^\ast\to VVV$ ($V=W,Z$) as two sensitive processes to probe $H$
at the LHC. To have large enough cross sections, we take the
semileptonic mode in the final states. We shall carefully analyze
the signal, irreducible background (IB), and all possible
reducible backgrounds (RBs), and impose a series of kinematic cuts
to effectively suppress the backgrounds. We shall see that the
heavy Higgs boson $H$ can be detected with reasonable integrated
luminosities at the 14 TeV LHC. Especially in the $pp\to
VH^\ast\to VVV$ process, a clear resonance peak of $H$ can be seen
experimentally.

Finally, we propose several physical observables from which the
anomalous coupling constants $f^{}_W$ and $f^{}_{WW}$ can be
measured experimentally. This provides a new high energy criterion
for new physics models beyond the SM. Only new physics models
giving $f^{}_W$ and $f^{}_{WW}$ consistently with the measured
values can survive, otherwise the models will be ruled out by this
new criterion. This helps us to find out the correct new physics
model reflecting the nature step by step.

This paper is organized as follows. Secs.\,II--IV are on studying
the formulation of the effective interactions and their
constraints. Secs.\,V--VIII are on the study of the LHC signatures
of $H$. In Sec.\,II. we present the details of the formulation of
the model-independent gauge and Yukawa couplings of $H$ in which
the anomalous gauge couplings are up to the dim-6 operators.
Sec.\,III is the study of the theoretical constraints on the
unknown coupling constants from the requirement of the unitarity
of the $S$-matrix. In Sec.\,IV, we study how the CMS $95\%$
exclusion limit on the SM Higgs boson leads to the lower bounds on
the unknown coupling constant. Combining the constraints given in
Secs.\,III and IV, we get the {\it available} range of the
anomalous coupling constants, with which $H$ is not excluded by
the yet known theoretical and experimental constraints. Sec.\,V is
a brief description of the general features of studying the LHC
signatures of $H$. In Sec.\,VI, we shall study the signal, IB, and
all the possible RB in weak-boson scattering, and we take proper
kinematic cuts for effectively suppressing the backgrounds from
analyzing the properties of the signal and backgrounds. Then we
show how the $M^{}_H=400,\,500$ and 800 GeV heavy neutral Higgs
boson can be detected at the 14 TeV LHC. Sec.\,VII is the study of
the $pp\to VH^\ast\to VVV$ process. We shall show that this
process is more sensitive than weak-boson scattering in the sense
that the resonance peak can be clearly seen, and the required
integrated luminosity is lower. In Sec.\,VIII, we shall show that
the anomalous coupling constants $f^{}_W$ and $f^{}_{WW}$ can be
measured by measuring both the cross section and certain
observable distributions of the final state particles. Sec.\,IX is
a concluding remark.

\section{Anomalous couplings of the non-standard heavy neutral Higgs
bosons}

For generality, we shall not specify the EW gauge group of the new
physics theories under consideration. The only requirement is that
the gauge group should contain an $SU(2)_L\times U(1)$ subgroup
with the gauge fields $W,Z$ and $\gamma$. Also, we shall not
specify the number of Higgs bosons and their group
representations, so that a Higgs boson in the Lagrangian may be
$SU(2)_L$ singlets, doublets, etc.

Let $\Phi^{}_1,\Phi^{}_2,\cdots\cdots$ be the original Higgs
fields (in various $SU(2)_L$ representations) in the Lagrangian.
The multi-Higgs potential $V(\Phi^{}_1,\Phi^{}_2,\cdots\cdots)$
will, in general, cause mixing between
$\Phi^{}_1,\Phi^{}_2,\cdots\cdots$ to form the mass eigenstates.
Let $\Phi^{}_h$ and $\Phi^{}_H$ be the lightest Higgs and a
heavier neutral heavy Higgs fields with Higgs bosons $h$ and $H$
(the neutral Higgs boson just heavier than $h$ and lighter than
other heavy Higgs bosons), respectively (gauge-phobic neutral
heavy Higgs bosons are not considered in this study). They are, in
general, mixtures of $\Phi^{}_1,\Phi^{}_2,\cdots\cdots$. So that
their vacuum expectation values (VEVs) $v^{}_h, v^{}_H$ are not
the same as the SM VEV $v$=246 GeV.

 In the following, we shall consider the
anomalous Yukawa couplings and anomalous gauge couplings
separately.

\subsection{Anomalous Yukawa Couplings}

The anomalous Yukawa couplings are relevant to our study of Higgs
decays. We are not interested in multi-Higgs-fermion couplings
which are irrelevant to our study.

As we have mentioned, we treat the 125--126 GeV Higgs boson $h$ as
SM-like, i.e., with negligible anomalous couplings. So that the
Yukawa couplings of $\Phi^{}_h$ to a fermion $f$ is
\bea                              %(1)
y^{h}_f\,\bar\psi_f \Phi^{}_h\psi_f, \label{Yh} \eea where
$y^{h}_f$ is the $\Phi^{}_h$-$f$-$\bar f$ Yukawa coupling constant
which is close to the SM Yukawa coupling constant $y^{SM}_f$.

For a NS heavy neutral Higgs boson $\phi^{}_H$, its Yukawa
coupling may not be the same as the standard Yukawa coupling. It
can  be seen that up to dim-6 operators, there is no new coupling
form other than the dim-4 Yukawa coupling contributing
\cite{Yukawa}. We thus formulate the anomalous Yukawa coupling of
$\Phi^{}_H$ to a fermion $f$ by
\begin{eqnarray}                              %(2)
          y^{H}_f\,\bar\psi_f \Phi^{}_H\psi_f\equiv C_f\,y^{SM}_f\,\bar\psi_f
          \Phi^{}_H\psi_f,
\label{C_f}
\end{eqnarray}
where $C_f$ is the anomalous factor of the Yukawa coupling. When
$C_f=1$, the coupling $y^H_f$ equals to the SM coupling $y^{}_f$.
In our study, the mostly relevant fermion is the $t$ quark since
$C_t$ concerns the $H$-$g$-$g$ vertex, i.e., the Higgs production
rate and the $H\to gg$ (Higgs decays to light hadrons) rate, and
the $H\to t\bar t$ decay rate as well.

The values of $C_t$ depends on the mixing between different
neutral Higgs bosons. So far there is no clear experimental
constraint on $C_t$. In the proposed new physics models, some of
the NS heavy neutral Higgs bosons has $C_t\approx 1$, while some
of the NS heavy neutral Higgs bosons have $C_t< 1$.

In our following studies, we consider both possibilities. {\it We
regard the $C_t\approx 1$ case as Type-I, and the $C_t<1$ case as
Type-II}.

Note that there are more than one Higgs bosons contributing to the
fermion mass $m^{}_f$, i.e.,
\bea                         %(3)
m^{}_f=\frac{1}{\sqrt 2}\left\{y^h_f v^{}_h+y^H_f
v^{}_H+\cdots\cdots\right\}. \label{m_f} \eea We know that, with
the SM Yukawa coupling $y^{SM}_f$ and $v=246$ GeV,
$m^{}_f=y^{SM}_f v/{\sqrt 2}$. Comparing this with (\ref{m_f}), we
obtain
\bea                        %(4)
\left\{\frac{y^h_f}{y^{SM}_f}\frac{
v^{}_h}{v}+\frac{y^H_f}{y^{SM}_f}\frac{v^{}_H}{v}+\cdots\cdots\right\}=1.
\label{m_f-c} \eea This serves as a constraint on the Yukawa
coupling constants and VEVs.

\subsection{Anomalous Gauge Couplings }

The effective gauge couplings of a Higgs boson in the multi-Higgs
system taking account of the Higgs mixing effect have not been
given in the published papers. We formulate them in the following.

We first consider the lightest Higgs boson $h$. Because of Higgs
mixing, the gauge coupling constant $g^{}_h$ of the lightest Higgs
field $\Phi^{}_h$ may not be the same as the $SU(2)_L$ gauge
coupling constant $g$. For a SM-like lightest Higgs boson,
$g^{}_h$ is close to $g$. With negligible anomalous couplings, the
dim-4 gauge couplings of the lightest Higgs field is
\bea                                          %(5)
\label{hcoupling}
&&{\cal L}^{(4)}_{hWW}=\frac{1}{2}g^2_hv^{}_h h
W_\mu W^\mu
\approx gM^{}_W\rho^{}_h h W_\mu W^\mu,\nonumber\\
&&{\cal L}^{(4)}_{hZZ}=\frac{1}{4c^2}g^2_hv^{}_h h Z_\mu Z^\mu
\approx \frac{gM^{}_W\rho^{}_h}{2c^2} h Z_\mu Z^\mu,\nonumber\\
&&\rho^{}_h\equiv\frac{g^2_h v^{}_h}{g^2 v},
%\label{hcoupling}
\eea                                        %
where $g$ is the $SU(2)_L$ gauge coupling, $v=246$ GeV, $M^{}_W$
is the $W$ boson mass, and $c\equiv\cos\theta^{}_W$.
%$V_\mu$ stands for the $W,Z$ bosons or the photon. %If the the

For the NS heavy neutral Higgs boson $H$, its gauge coupling
$g^{}_H$ may not be close to $g$ due to the Higgs mixing depending
on the property of $H$. Similar to (\ref{hcoupling}), the dim-4
gauge coupling of $H$ is
\bea                             %(6)
\label{dim-4Hcoupling} &&{\cal
L}^{(4)}_{HWW}=\frac{1}{2}g^2_Hv^{}_H H W_\mu W^\mu
\approx gM^{}_W\rho^{}_H H W_\mu W^\mu,\nonumber\\
&&{\cal L}^{(4)}_{HZZ}=\frac{1}{4c^2}g^2_Hv^{}_H H Z_\mu Z^\mu
\approx \frac{gM^{}_W\rho^{}_H}{2c^2} H Z_\mu Z^\mu,\nonumber\\
&&\rho^{}_H\equiv\frac{g^2_H v^{}_H}{g^2 v},
\eea                               %
{\it Eq.\,(\ref{dim-4Hcoupling}) differs from the SM form only by
an extra factor $\rho^{}_H$, i.e.,  $g^2 v\Longrightarrow g^2 v
\rho^{}_H$}. Since $\rho^{}_H$ depends on the specific mixing
between $H$ and other Higgs bosons, we take it as an unknown
parameter here.

 Beyond the dim-4 coupling
(\ref{dim-4Hcoupling}), there can also be dim-6 anomalous gauge
couplings of $H$. The form of the dim-6 anomalous gauge couplings
for a single-Higgs system (with the dim-4 coupling the same as the
SM interaction) was given in
Refs.\,\cite{Hagiwara}\cite{Buchmuller} and a detailed review of
this was given in Ref.\,\cite{G_G}. Now we are dealing with a
multi-Higgs system with the dim-4 coupling shown in Eq.\,
(\ref{dim-4Hcoupling}). Referring to the relation between the
dim-4 and dim-6 couplings given in
Refs.\,\cite{Hagiwara}\cite{G_G}, we write down our dim-6
couplings as
%It is of the form
% \cite{HVV03,HVV09}
\begin{equation}                    %(7)
{\cal L}^{(6)}_{H VV} ~\,=~\, \sum_n \frac{f_n}{\Lambda^2} {\cal
O}_n \,. \label{Leff}
\end{equation}
where $\Lambda$ is the scale below which the effective Lagrangian
holds. When it is needed to specify the value of $\Lambda$ in some
cases, we shall take $\Lambda$=3 TeV as an example.
 The gauge-invariant dimension-6 operators ${\cal
O}_n$'s are

\begin{widetext}

\begin{eqnarray}                    %(8)
&&\hspace{-0.5cm} {\cal O}_{BW} =  \Phi^{\dagger}_H \hat{B}_{\mu
\nu}
\hat{W}^{\mu \nu} \Phi_H, %\nonumber \\
%&&\hspace{-0.5cm}
~~~~{\cal O}_{DW} = \mbox{Tr}([D_{\mu},\hat{W}_{\nu\rho}],[D^{\mu},\hat{W}^{\nu\rho}]),%\nonumber \\
%&&\hspace{-0.5cm}
~~~~{\cal O}_{DB}=-\frac{{g^\prime}^2}{2} (\partial_\mu
B_{\nu\rho}) (\partial^\mu B^{\nu\rho}),\nonumber \\
&&\hspace{-0.5cm} {\cal O}_{\Phi,1} =  (D_\mu \Phi^{}_H)^\dagger
\Phi^\dagger_H \Phi^{}_H
(D^\mu \Phi^{}_H), %\nonumber \\
%&&\hspace{-0.5cm}
~~~~{\cal O}_{\Phi,2} =\frac{1}{2}
\partial^\mu\kc{\Phi^\dagger_H \Phi^{}_H}
\partial_\mu\kc{\Phi^\dagger_H \Phi^{}_H},~~%\nonumber \\
{\cal O}_{\Phi,3} =\frac{1}{3} (\Phi^\dagger_H \Phi^{}_H)^3,\nonumber\\
&&\hspace{-0.5cm} {\cal O}_{WWW}=\mbox{Tr}[\hat{W}_{\mu
\nu}\hat{W}^{\nu\rho}\hat{W}_{\rho}^{\mu}]
,~~%\nonumber \\
{\cal O}_{WW} = \Phi^{\dagger}_H \hat{W}_{\mu \nu}
\hat{W}^{\mu \nu} \Phi^{}_H , %\nonumber \\
%&&\hspace{-0.5cm}
~~~~{\cal O}_{BB} = \Phi^{\dagger}_H \hat{B}_{\mu \nu}
\hat{B}^{\mu
\nu} \Phi^{}_H , \nonumber \\
&&\hspace{-0.5cm}{\cal O}_W  = (D_{\mu} \Phi^{}_H)^{\dagger}
\hat{W}^{\mu \nu}  (D_{\nu} \Phi^{}_H),%\nonumber \\
%&&\hspace{-0.5cm}
~~~~{\cal O}_B  =  (D_{\mu} \Phi^{}_H)^{\dagger} \hat{B}^{\mu \nu}
(D_{\nu} \Phi^{}_H), \label{O}
\end{eqnarray}
\end{widetext}
where $\hat B_{\mu\nu}$ and $\hat W_{\mu\nu}$ stand for
\begin{eqnarray}                       %(9)
\hat{B}_{\mu \nu} = i \frac{g'_H}{2} B_{\mu \nu},\;\;\;\;\;\;\;\;
\hat{W}_{\mu \nu} = i \frac{g^{}_H}{2} \sigma^a W^a_{\mu \nu},
\label{B,W}
\end{eqnarray}
%\end{widetext}
in which $g^{}_H$ and $g^\prime_H$ are the $SU(2)_L$ and $U(1)$
gauge coupling constants  of $H$, respectively. It has been shown
that the operators ${\cal O}_{\Phi,1}$, ${\cal O}_{BW}$, ${\cal
O}_{DW}$, ${\cal O}_{DB}$ are related to the two-point functions
of the weak bosons, so that they are severely constrained by the
precision EW data \cite{G_G}. For example, ${\cal O}_{BW}$ and
${\cal O}_{\Phi,1}$ are related to the oblique correction
parameters $S$ and $T$, and are thus strongly constrained by the
precision EW data. The $2\sigma$ constraints on
$|f_{BW}/\Lambda^2|$ and $|f_{\Phi,1}/\Lambda^2|$ are:
$|f_{BW}/\Lambda^2|, |f_{\Phi,1}/\Lambda^2|<
O(10^{-2})$~TeV$^{-2}$ \cite{ZKHY03}. The operators ${\cal
O}_{\Phi,2}$ and ${\cal O}_{\Phi,3}$ are related to the triple and
quartic Higgs boson self-interactions, and have been studied in
detail in Ref.~\cite{BHLMZ}. The operator $O_{WWW}$ is related to
the weak-boson self-couplings, so that it is irrelevant to the
present study. Furthermore, the ATLAS and CMS experiments on
testing the triple gauge couplings \cite{3gauge} show stronger and
stronger constraints on the anomalous triple gauge coupling. So
that we ignore the operator $f^{}_{WWW}{\cal O}_{WWW}/\Lambda^2$
in our present study. The precision and low energy EW data are not
sensitive to the remaining four operators ${\cal O}_{WW}$, ${\cal
O}_{BB}$, ${\cal O}_{W}$, and ${\cal O}_{B}$, so these four
operators are what we shall pay special attention in our study in
high energy processes.

The relevant effective Lagrangian expressed in terms of the photon
field $A_\mu$, the weak-boson fields $W^\pm_\mu$, $Z_\mu$, and the
Higgs boson field $H$ is
\begin{widetext}

\begin{eqnarray}                              %(10)
\hspace{-0.4cm}{\cal
L}^{(6)}_{HVV}&=&g^{}_{H\gamma\gamma}HA_{\mu\nu}A^{\mu\nu}
+g^{(1)}_{HZ\gamma}A_{\mu\nu}Z^\mu\partial^\nu H~~~~%\nonumber\\
%\hspace{0.4cm}
+g^{(2)}_{HZ\gamma}HA_{\mu\nu}Z^{\mu\nu}
+g^{(1)}_{HZZ}Z_{\mu\nu}Z^\mu\partial^\nu H~~~~%\nonumber\\
%\hspace{0.4cm}
+g^{(2)}_{HZZ}HZ_{\mu\nu}Z^{\mu\nu}\nonumber\\
&&+g^{(1)}_{HWW}(W^+_{\mu\nu} W^{-\mu}\partial^\nu H+{\rm h.c.})~~~~%\nonumber\\
%\hspace{0.4cm}
+g^{(2)}_{HWW}HW^+_{\mu\nu}W^{-\mu\nu}, \label{LHeff}
\end{eqnarray}
\end{widetext}
%As we mentioned below Eq.\,(\ref{dim-4Hcoupling}) that the present
%${\cal L}^{(4)}_{HVV}$ differs from the SM form only by an extra
%factor $\rho^{}_H$.
and the anomalous couplings $g^{(i)}_{HVV}$ with~$i=1,2$ in our
case are related to the anomalous couplings $f_n$'s by
\begin{widetext}

\begin{eqnarray}                           %(11)
&&\hspace{-0.4cm}g^{}_{H\gamma\gamma}=-gM^{}_W\rho^{}_H\frac{s^2(f_{BB}
+f_{WW})}{2\Lambda^2},%\nonumber\\
%&&\hspace{-0.4cm}
~~~~g^{(1)}_{HZ\gamma}=gM^{}_W\rho^{}_H\frac{s(f_W-f_B)}{2c\Lambda^2},
%\nonumber\\
%&&
~~~~g^{(2)}_{HZ\gamma}=gM^{}_W\rho^{}_H\frac{s[s^2f_{BB}
-c^2f_{WW}]}{c\Lambda^2},\nonumber\\
&&\hspace{-0.4cm}g^{(1)}_{HZZ}=gM^{}_W\rho^{}_H\frac{c^2f_W+s^2f_B}{2c^2\Lambda^2},
%\nonumber\\
%&&
~~g^{(2)}_{HZZ}=-gM^{}_W\rho^{}_H\frac{s^4f_{BB}
+c^4f_{WW}}{2c^2\Lambda^2},\nonumber\\
&&\hspace{-0.4cm}g^{(1)}_{HWW}=gM^{}_W\rho^{}_H\frac{f_W}{2\Lambda^2},%\nonumber\\
%&&
~~~~~~~~~~~~~g^{(2)}_{HWW}=-gM^{}_W\rho^{}_H\frac{f^{}_{WW}}{\Lambda^2},
\label{g} \label{g(f)}
\end{eqnarray}
\end{widetext}
in which $s\equiv \sin\theta_W,~c\equiv \cos\theta_W$. These
formulas are similar to those given in Ref.\,\cite{G_G} but with
an extra factor $\rho^{}_H$ reflecting the Higgs mixing effect in
the overall constant.% associated with every anomalous coupling
%constant $f^{}_n$.

So including the dim-4 and dim-6 anomalous couplings, there are
altogether five new parameters, namely $\rho^{}_H, f^{}_W,
f^{}_{WW}, f^{}_B$ and $f^{}_{BB}$. We see from Eq.\,(\ref{g(f)})
that {\it the parameters $f^{}_B$ and $f^{}_{BB}$ are not related
to the $HWW$ couplings. They appear in the $HZZ$ couplings with
the small factors $s^2$ and $s^4$, respectively. They mainly
contribute to the $H\gamma\gamma$ and $HZ\gamma$ couplings.}

The $HVV$ operators in (\ref{LHeff}) contain extra derivatives
relative to (\ref{dim-4Hcoupling}). So that ${\cal L}^{(6)}_{HVV}$
is momentum dependent in the momentum representation, i.e., the
dim-6 coupling has an extra factor of
 $O(k^2/\Lambda^2)$ relative to the dim-4 coupling. This means that the effect of ${\cal L}^{(6)}_{HVV}$ is small in
the low momentum region but it is enhanced in high energy
processes. This is the reason why we take into account both ${\cal
L}^{(4)}_{HVV}$ and ${\cal L}^{(6)}_{HVV}$ in our study.

 To see the details of the
momentum dependence, we list, in the following, the momentum
representations of the $HVV$ interactions in (\ref{LHeff}).

\begin{figure}[h]                                 %Fig.1
\centering   \bc
\includegraphics[width=5cm,height=3cm]{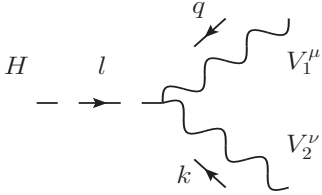}
\ec \vspace{-0.8cm} \caption{Illustration of the momenta in the
$HVV$ interactions in Eq.\,(\ref{LHeff}).} \label{HVV}
\end{figure}

The three momenta in the $HVV$ vertices in (\ref{LHeff}) are
illustrated in FIG.\,\ref{HVV}, in which $l$ stands for the
momentum of $H$, $q$ and $k$ stand for the momenta of the two
gauge fields $V_{1}^\mu$ and $V_{2}^\nu$, respectively. They
satisfy
\bea                             %(12)
l_\mu+q_\mu+k_\mu=0. \label{mom-cons}
\eea

\noindent {\it (a) The $H\gamma\gamma$ Interactions}
\bea                                %(13)
g^{}_{H\gamma\gamma}HA_{\mu\nu}A^{\mu\nu}&\to&
2g^{}_{H\gamma\gamma}(q_\nu k_\mu-g_{\mu\nu}q\cdot k)A^\mu A^\nu
H\nonumber\\
&=&-2gM^{}_W\rho^{}_H\frac{s^2(f_{BB}
+f_{WW})}{2\Lambda^2}\nonumber\\
&&\times (q_\nu k_\mu-g_{\mu\nu}q\cdot k)A^\mu A^\nu H
\label{Hgammagamma} \eea

\null\vspace{0.4cm}
\noindent {\it (b) The $HZ\gamma$ Interactions}

Taking $V_1^\mu=A^\mu,V_2^\mu=Z^\mu$, we have
\bea                             %(14)
\label{csHZgamma}
&&\hspace{-0.4cm}g_{HZ\gamma}^{(1)}A_{\mu\nu}Z^\mu\partial^{\nu}H+g_{HZ\gamma}^{(2)}HA_{\mu\nu}Z^{\mu\nu}\nonumber\\
&&\hspace{-0.4cm}\to \bigg[ g_{HZ\gamma}^{(1)} ( q_\mu q_\nu -
q^2g_{\mu\nu}
+ q_\nu k_\mu - g_{\mu\nu}q\cdot k )\nonumber\\
&&\hspace{-0.0cm}+ 2g_{HZ\gamma}^{(2)} ( q_\nu k_\mu - g_{\mu\nu}q\cdot k )\bigg]A^\mu Z^\nu H\nonumber\\
%&&\hspace{-0.4cm}=\frac{gM_W\rho^{}_H}{c\Lambda^2}\bigg[\frac{1}{2}s(f^{}_W-f^{}_B)(q^{}_\mu
%q^{}_\nu-q^2 g^{}_{\mu\nu}+q^{}_\nu k^{}_\mu%\nonumber\\
%%&&\hspace{-0.4cm}
%-g^{}_{\mu\nu}q\cdot k )\nonumber\\
%&&\hspace{-0.0cm}+2s(s^2f^{}_{BB}-c^2 f^{}_{WW})%\nonumber\\
%%&&~~~~
%( q_\nu k_\mu - g_{\mu\nu}q\cdot k )\bigg]A^\mu Z^\nu
%H\nonumber\\
&&\hspace{-0.4cm}=\frac{gM_W\rho^{}_Hs}{2c\Lambda^2}\bigg[(f^{}_W-f^{}_B)(q^{}_\mu
q^{}_\nu-q^2 g^{}_{\mu\nu})
+\bigg(f^{}_W-f^{}_B\nonumber\\
&&\hspace{-0.2cm}+4(s^2f^{}_{BB}-c^2 f^{}_{WW})\bigg)( q_\nu k_\mu
- g_{\mu\nu}q\cdot k )\bigg]A^\mu Z^\nu H
\label{csHZgamma}                          %
\eea                                  %
Neglecting the small term proportional to ${s}^2$, we have
\bea                                         %(15)
&&\hspace{-0.6cm}g_{HZ\gamma}^{(1)}A_{\mu\nu}Z^\mu\partial^{\nu}H+g_{HZ\gamma}^{(2)}HA_{\mu\nu}Z^{\mu\nu}\nonumber\\
 &&\hspace{-0.4cm}\approx
\frac{gM_W\rho^{}_Hs}{2c\Lambda^2}\bigg[(f^{}_W-f^{}_B)(q^{}_\mu
q^{}_\nu-q^2 g^{}_{\mu\nu})\nonumber\\
&&\hspace{-0.4cm}+\bigg(f^{}_W-f^{}_B-4f^{}_{WW} \bigg)( q_\nu
k_\mu - g_{\mu\nu}q\cdot k )\bigg]A^\mu Z^\nu H.
\label{HZgamma}                                     %
\eea                                      %

\null\noindent {\it (c) The $HWW$ Interactions}
\bea                      %(16)
&&g_{HWW}^{(1)}(W^+_{\mu\nu}W^{-\mu}\partial^\nu H+{\rm h.c.})+g_{HWW}^{(2)}HW^+_{\mu\nu}W^{-\mu\nu}\nonumber\\
&&~~\to  \bigg[g_{HWW}^{(1)}(q_\mu q_\nu - q^2g_{\mu\nu} + k_\mu
k_\nu - k^2g_{\mu\nu})\nonumber\\
&&~~+ 2(g_{HWW}^{(1)}+g_{HWW}^{(2)})(q_\nu k_\mu - q\cdot
kg_{\mu\nu})\bigg]W^{+\mu}W^{-\nu}H\nonumber\\
&&~~=\frac{gM_W\rho^{}_H}{2\Lambda^2}\bigg[f^{}_W(q_\mu q_\nu -
q^2g_{\mu\nu} + k_\mu k_\nu - k^2g_{\mu\nu})\nonumber\\
&&~~+2(f^{}_W-2f^{}_{WW})(q_\nu k_\mu - q\cdot
kg_{\mu\nu})\bigg]W^{+\mu}W^{-\nu}H \label{HWW} \eea

\null\noindent{\it (d) The $HZZ$ Interactions}
\bea                           %(17)
&&\hspace{-0.4cm}g_{HZZ}^{(1)}Z_{\mu\nu}Z^\mu\partial^{\nu}H+g_{HZZ}^{(2)}HZ_{\mu\nu}Z^{\mu\nu}\nonumber\\
  &&\hspace{-0.2cm}  \to
 \bigg[\frac{1}{2}g_{HZZ}^{(1)}( q_\mu q_\nu -
q^2g_{\mu\nu}+k_\mu k_\nu - k^2g_{\mu\nu}
    )\nonumber\\
&&\hspace{-0.2cm} + ( g_{HZZ}^{(1)} + 2g_{HZZ}^{(2)} )( q_\nu
k_\mu -
g_{\mu\nu}q\cdot k ) \bigg] Z^\mu Z^\nu H\nonumber\\
&&\hspace{-0.2cm}=\frac{gM_W\rho^{}_H}{2c^2\Lambda^2}\bigg[\frac{1}{2}(c^2f^{}_W+s^2f^{}_B)\nonumber\\
&&\hspace{-0.2cm}\times (q_\mu q_\nu - q^2g_{\mu\nu}+k_\mu k_\nu -
k^2g_{\mu\nu})\nonumber\\
&&\hspace{-0.2cm}+(c^2f^{}_W+s^2f^{}_B-2s^4f^{}_{BB}-2c^4f^{}_{WW})\nonumber\\&&\hspace{-0.2cm}
\times (q_\nu k_\mu - g_{\mu\nu}q\cdot k ) \bigg] Z^\mu Z^\nu H
\label{csHZZ}                               %
\eea
Neglecting the small terms proportional to ${s}^2$ and ${s}^4$, we have                                       %
\bea                               %(18)
&&\hspace{-0.4cm}g_{HZZ}^{(1)}Z_{\mu\nu}Z^\mu\partial^{\nu}H+g_{HZZ}^{(2)}HZ_{\mu\nu}Z^{\mu\nu}\nonumber\\
 &&\hspace{-0.2cm}\approx
\frac{gM_W\rho^{}_H}{2c^2\Lambda^2}\bigg[\frac{1}{2}f^{}_W( q_\mu
q_\nu - q^2g_{\mu\nu}+k_\mu k_\nu - k^2g_{\mu\nu})\nonumber\\
&&\hspace{-0.2cm}+(f^{}_W-2f^{}_{WW})( q_\nu k_\mu -
g_{\mu\nu}q\cdot k ) \bigg]Z^\mu Z^\nu H.
 \label{HZZ}                                      %
 \eea                                            %

 Now the gauge boson masses, especially the $W$ boson mass, are also
contributed by more than one Higgs fields. Since ${\cal
L}^{(6)}_{HVV}$ contains extra derivatives, it does not contribute
to the $W$ boson mass. From (\ref{hcoupling}) and
(\ref{dim-4Hcoupling}) we see that
\bea                             %(19)
M_W^2&=&\frac{1}{4}(g_h^2v_h^2+g_H^2v_H^2+\cdots\cdots)\nonumber\\
&=&\frac{1}{4}g^2v(\rho_hv_h+\rho_Hv_H+\cdots\cdots).
\label{Wmass} \eea Comparing with the SM $W$ boson mass
$M_W^2=g^2v^2/4$, we obtain
\bea                            %(20)
\rho_h\frac{v_h}{v}+\rho_H\frac{v_H}{v}+\cdots\cdots=1.
\label{Wmass-c} \eea This serves as another constraint on the
gauge coupling constants and VEVs. It is easy to see that the two
constraints (\ref{Wmass-c}) and (\ref{m_f-c}) can be satisfied
simultaneously.

\section{Unitarity Constraints on the Anomalous Coupling Constants}

As we mentioned in the last section, the anomalous interactions in
${\cal L}^{(4)}_{HVV}+{\cal L}^{(6)}_{HVV}$ include five unknown
anomalous coupling constants $\rho_H, f^{}_W, f^{}_{WW}, f^{}_B,$
and $f^{}_{BB}$. Low energy observables are insensitive to the
related operators in ${\cal L}^{(6)}_{HVV}$. We are going to study
certain constraints from high energy processes. In this section,
we study the theoretical constraint obtained from the requirement
of the unitarity of the $S$-matrix. In the next section, we shall
study the experimental constraint obtained from the CMS $95\%$ CL
exclusion bound on the SM Higgs boson.

We would like to emphasize that we are not aiming at precision
calculations in this and the next sections. Instead, our purpose
is to find out a rough range of the anomalous coupling constants
$f^{}_W$ and $f^{}_{WW}$ with $f^{}_W$ and $f^{}_{WW}$ inside
which the heavy Higgs boson is not excluded by the existing
theoretical and experimental constraints, so that the study of
probing the heavy Higgs boson at 14 TeV LHC makes sense.

Since the operators in ${\cal L}^{(6)}_{HVV}$ are momentum
dependent, it will violate the unitarity of the $S$-matrix at high
energies (note that the CM energy can not exceed $\Lambda$ in the
effective Lagrangian theory). So that the requirement of the
unitarity of the $S$-matrix can give constraints on the size of
the anomalous coupling constants. This kind of study has been
given in several papers \cite{unitarity} in which the effective
couplings for the
single-Higgs system was taken, %dim-4 interaction is taken to be
%the SM form,
and the study is a single-parameter analysis. We cannot simply
take such constraint in our study because we are studying the
effective couplings in a multi-Higgs system taking account of the
contributions of both the lightest SM-like Higgs $h$ and a heavier
neutral Higgs boson $H$ with $\rho^{}_h,\rho^{}_H\ne 1$. In the
following, to get the order of magnitude constraints, we study the
unitarity constraints for our case in the effective $W$
approximation (EWA).

The strongest constraints come from the longitudinal weak-boson
scattering since the polarization vector $\epsilon_L^\mu$ of $W_L$
($Z_L$) contains extra momentum dependence. To the precision of
EWA, it is reasonable to neglect the small terms of $O(s^2)$ and
$O(s^4)$ in the anomalous $HZZ$ coupling as in the last step in
Eq.\,(\ref{HZZ}). Then we see from (\ref{HWW}) and (\ref{HZZ})
that the relevant anomalous $HWW$ and $HZZ$ couplings contain only
three unknown coupling constants $\rho^{}_H$, $f^{}_W$ and
$f^{}_{WW}$, irrelevant to $f^{}_B$ and $f^{}_{BB}$.

Expressing the $S$-matrix by $S=1-iT$, the unitarity of the
$S$-matrix reads
\bea                           %(21)
|S^\dagger S|=|1-iT|^2=1 \label{Sunitarity} \eea which leads to
the following requirement
\bea                          %(22)
%&&T^\dag T+i(T-T^\dag)=0\nonumber\\
%\langle a|T^\dag T|a\rangle-2{\rm Im}\langle a |T|a\rangle=0\nonumber\\
&&\hspace{-0.8cm}({\rm Re}\langle a|T|a\rangle)^2+ ({\rm
Im}\langle a|T|a\rangle-1)^2+\sum_{
|b\rangle\ne|a\rangle}\left|\langle
b|T|a\rangle\right|^2=1\nonumber\\
&&\Longrightarrow ({\rm Re}\langle a|T|a\rangle)^2+ \sum_{
|b\rangle\ne|a\rangle}\left|\langle b|T|a\rangle\right|^2\le 1.
\label{optical_theorem}
\eea

When we take $|a\rangle=|W_LW_L\rangle$, the leading final state
$\langle b|$ is $\langle W_LW_L|$ and $\langle Z_LZ_L|$. In
certain regions of the anomalous coupling constants, the leading
matrix element may be small, so that other non-leading final
states should also be considered. Thus we also include $\langle
b|=\langle W_TW_T|$, and $\langle Z_TZ_T|$. Similarly, when we
take $|a\rangle=|Z_LZ_L\rangle$, we take $\langle b|=\langle
Z_LZ_L|,\,\langle W_TW_T|$, and $\langle Z_TZ_T|$.

As usual, the unitarity constraints is to be calculated in the
partial wave expression which was studied in detail in
Ref.\,\cite{JW}. It is well-known that the $S$-wave contribution
is dominant. So we only calculate the matrix elements of the
$S$-wave amplitude $T^0_{}$. For $|a\rangle=|W_LW_L\rangle$ and
$|a\rangle=|Z_LZ_L\rangle$, the unitarity constraints read
\bea                                  %(23)
&&\hspace{-0.4cm}|{\rm Re}\langle W_L^+W_L^-|T^0|W_L^+W_L^-\rangle|^2+|\langle Z_LZ_L|T^0|W_L^+W_L^-\rangle|^2\nonumber\\
    &&\hspace{-0.4cm}+2|\langle W_+^+W_+^-|T^0|W_L^+W_L^-\rangle|^2+2|\langle Z_+Z_+|T^0|W_L^+W_L^-\rangle|^2\nonumber\\
    &&\le 1,
\label{W_LW_L-VV-UC} \eea
\null\noindent and
\bea                                 %(24)
&&|{\rm Re}\langle Z_LZ_L|T^0|Z_LZ_L\rangle|^2+2|\langle Z_\pm Z_\pm|T^0|Z_LZ_L\rangle|^2\nonumber\\
&&~~+|\langle W_L^+ W_L^-|T^0|Z_LZ_L\rangle|^2+2|\langle
W_\pm^+W_\pm^-|T^0|Z_LZ_L\rangle|^2\nonumber\\
&&~~\le 1.
\label{Z_LZ_L-VV-UC}                  %
\eea                                    %

 In our study, we have taken into account the contributions of
 both $h$ and $H$. These kinds of results have not been given in the
 published papers. We shall present our analytical results and
 numerical analysis as follows. We give the results in the center-of-mass (c.m.) frame, and
express the scattering amplitudes in terms of the {\sf s,t,u}
parameters.

\begin{widetext}

\subsection{\texorpdfstring{\bm{$W_L^+W_L^-\rightarrow VV$}}{Longitudinal W Bosons Scattering}}

\null\vspace{-1cm}
\bea                                 %(25)
&&{\rm Re}\langle
W_L^+W_L^-|T^0|W_L^+W_L^-\rangle=-\frac{g^2}{64\pi}%~~~~~~~~\nonumber\\
%&&\hspace{0.2cm}\times
\left\{\left[\rho_H^2(1-\frac{M_W^2}{\Lambda^2}f^{}_W)^2+\rho_h^2-1\right]\frac{{\sf
s}}{M_W^2}+O({\sf s}^0)\right\}.~~~~~~~~
\label{TW_LW_L-W_LW_L}                                %
\eea
\bea                                          %(26)
&&\langle
Z_LZ_L|T^0|W_L^+W_L^-\rangle=\frac{g^2}{32\pi}\bigg\{\bigg[\rho_H^2\bigg(1-\frac{M_W^2}{\Lambda^2}f^{}_W\bigg)%\nonumber\\
%&&~~\times
\bigg(\frac{M_Z^2}{\Lambda^2}f^{}_W-1\bigg)-\rho_h^2+1\bigg]\frac{{\sf
s}}{M_W^2}+O({\sf s}^0)\bigg\}.
\label{TW_LW_L-Z_LZ_L}                          %
\eea                                   %
\bea                                          %(27)
\langle
W_\pm^+W_\pm^-|T^0|W_L^+W_L^-\rangle=\frac{\rho_H^2g^2}{32\pi}\bigg(1-\frac{M_W^2}{\Lambda^2}f^{}_W\bigg)%\nonumber\\
%&&\hspace{0.4cm}\times
(2f^{}_{WW}-f_W)\frac{{\sf s}}{\Lambda^2}+O({\sf s}^0).
\label{TW_LW_L-W_+W_+}
\eea                                %
\bea                                     %(28)
\langle Z_\pm
Z_\pm|T^0|W_L^+W_L^-\rangle&=&\displaystyle\frac{\rho_H^2g^2}{32\pi
}\bigg(1-\frac{M_W^2}{\Lambda^2}f^{}_W\bigg)%\nonumber\\
%&&\times
(2f^{}_{WW}-f^{}_W)\frac{{\sf s}}{\Lambda^2}+O({\sf s}^0).
\label{TW_LW_L-Z_+Z_+}                      %
\eea                                          %

In (\ref{TW_LW_L-W_LW_L})--(\ref{TW_LW_L-Z_+Z_+}), the terms with
$\rho^{}_H$ are the contributions of $H$, and those with
$\rho^{}_h$ are contributions of $h$. We see from
(\ref{TW_LW_L-W_+W_+}) and (\ref{TW_LW_L-Z_+Z_+}) that, in these
two matrix elements, the leading terms contain only the
contributions of $H$ (from its dim-6 couplings).

%\null\vspace{-0.53cm}

\subsection{\texorpdfstring{\bm{$Z_LZ_L\to VV$}}{Longitudinal Z Bosons Scattering}}

Since there are all ${\sf s}, {\sf t}$ and ${\sf u}$ channel
contributions in $Z_LZ_L\to Z_LZ_L$, the leading $O({\sf s}^1)$
terms in the three channels just cancel with each other. So that
\bea                                   %(29)
{\rm Re}\langle Z_LZ_L|T^0|Z_LZ_L\rangle=O({\sf s}^0).
\label{TZ_LZ_L}                            %
\eea                                     %

Results of other final states are

\bea                                   %(30)
\langle Z_\pm Z_\pm|T^0|Z_LZ_L\rangle=\frac{\rho_H^2g^2}{32\pi
}\bigg\{\left(1-\frac{M_Z^2}{\Lambda^2}f^{}_W\right)%\nonumber\\
%&&~~~~\times
(2f^{}_{WW}-f^{}_W)\frac{{\sf s}}{\Lambda^2}+O({\sf
s}^0)\bigg\},
\label{TZ_LZ_L-Z_+Z_+}                        %
\eea                                           %
and
\bea                                   %(31)
\langle
W_\pm^+W_\pm^-|T^0|Z_LZ_L\rangle=\frac{\rho_H^2g^2}{32\pi}\bigg\{(2f^{}_{WW}-f^{}_W)%\nonumber\\
%&&~~~~\times
\left(1-\frac{M_Z^2}{\Lambda^2}f^{}_W\right)\frac{{\sf
s}}{\Lambda^2}+O({\sf s}^0)\bigg\}.
\label{TZ_LZ_L-W^+W^+}                            %
\eea                                              %

We se that in (\ref{TZ_LZ_L})--(\ref{TZ_LZ_L-W^+W^+}), all the
leading terms contain only the contributions of $H$ (from its dim-6 couplings).\\

\end{widetext}

\begin{widetext}

\begin{figure}[h]                                 %Fig.2
\centering   \bc
\hspace{0.5cm}\includegraphics[width=15.5cm,height=6.5cm]{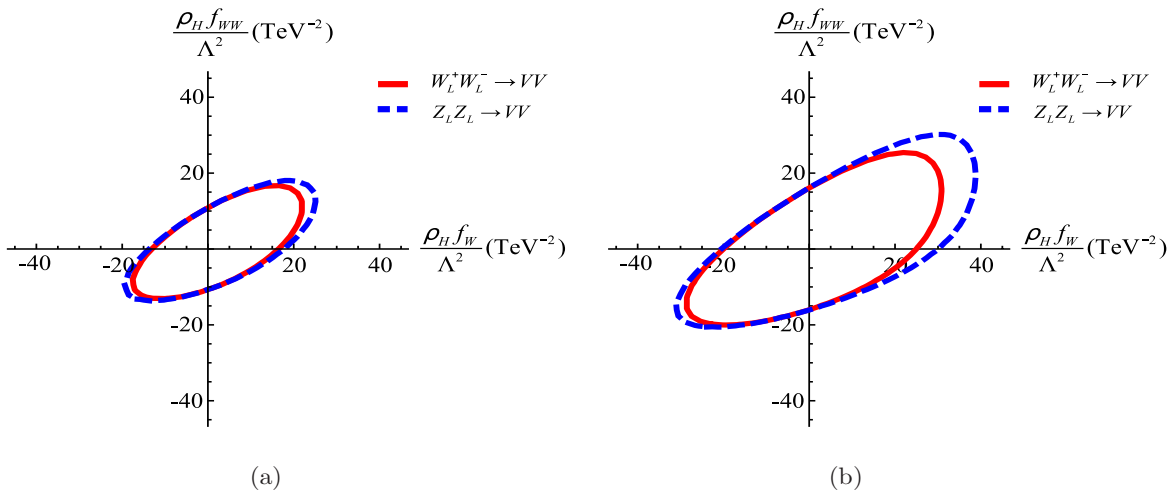}
\vspace{0.4cm}
 \ec
\vspace{-0.7cm} \caption{Unitarity bounds on $f^{}_W$ and
$f^{}_{WW}$ in which (a) is with $\rho^{}_h=0.8$ and
$\rho^{}_H=0.6$; (b) is with $\rho^{}_h=0.9$ and $\rho^{}_H=0.4$.
The red and blue-dashed contours are boundaries of the allowed
regions obtained from $W_L^+W_L^-\to VV$
[Eq.\,(\ref{W_LW_L-VV-UC})] and $Z_LZ_L\to VV$
[Eq.\,(\ref{Z_LZ_L-VV-UC})], respectively.}
\label{UB}                                %
\end{figure}                        %
\end{widetext}

With all the above results, we are ready to analyze the unitarity
constraints on the anomalous coupling constants $f^{}_W$ and
$f^{}_{WW}$. Since we are interested in weak-boson scattering at
high energies in which ${\cal L}^{(6)}_{HVV}$ is enhanced, we
shall only keep the terms with leading power of ${\sf s}$ in all
the above results. In our numerical analysis, we simply take the
{\sf s} parameter to be its highest value ${\sf s}=\Lambda^2$. We
shall study such constraints numerically performing a
two-parameter analysis. Before doing that, we need to specify the
other unknown parameters. First of all, as we have mentioned in
Sec.\,II, we shall take $\Lambda=3$ TeV as an example. For
$\rho^{}_h$, the known SM-like properties of $h$ means that
$\rho^{}_h$ should not be so different from 1. We shall take
$\rho^{}_h=0.8,\,0.9$ as two examples. For $\rho^{}_H$, we shall
see in the next section that if $\rho^{}_H>0.6$, the heavy neutral
Higgs boson $H$ can hardly avoid being excluded by the CMS $95\%$
CL exclusion bounds on the SM Higgs boson. Therefore, for an
existing $H$, $\rho^{}_H$ should be less than 0.6. We shall take
$\rho^{}_H=0.6,\,0.4$ as two examples. The results of our analysis
are shown in FIG.\,\ref{UB} in which FIG.\,\ref{UB}(a) is with
$\rho^{}_h=0.8$, $\rho^{}_H=0.6$, and FIG.\,\ref{UB}(b) is with
$\rho^{}_h=0.9$, $\rho^{}_H=0.4$. In FIG.\,\ref{UB}, the red and
blue-dashed contours are boundaries of the allowed regions
obtained from $W_L^+W_L^-\to VV$ [Eq.\,(\ref{W_LW_L-VV-UC})] and
$Z_LZ_L\to VV$ [Eq.\,(\ref{Z_LZ_L-VV-UC})], respectively.

We see that $\rho^{}_H f^{}_W/\Lambda^2$ and $\rho^{}_H
f^{}_{WW}/\Lambda^2$ are constrained up to a few tens of
TeV$^{-2}$ which is different from the results given in
Ref.\,\cite{unitarity}.

So far we have not concerned the unitarity bounds on $f^{}_B$ and
$f^{}_{BB}$. In principle, they can be obtained by studying the
scattering processes $W_LW_L\to \gamma\gamma$ and $W_LW_L\to
Z\gamma$. However, since the photon has only transverse
polarizations, such bounds will be weaker. Actually, in the next
section, we shall argue that we may make the approximation of
neglecting the anomalous coupling constants in the dim-6 couplings
of the $H\gamma\gamma$ and $HZ\gamma$ couplings.

\section{Experimental Constraints on Anomalous Coupling Constants}

After the discovery of the 125--126 GeV Higgs boson in 2012, the
CMS (ATLAS) Collaboration has made a lot of measurements on
excluding the SM Higgs boson with mass up to 1 TeV (600 GeV)
\cite{CMS_HIG_13_002}\cite{ATLAS_CMS12}\cite{others} at $95\%$
C.L. For a NS heavy neutral Higgs boson, it must have large enough
anomalous couplings to reduce its production cross section to
avoid being excluded by the CMS experiments. This provides the
possibility of constraining the anomalous coupling constants
experimentally. In this section, we study such experimental
bounds. Values of the anomalous coupling constants consistent with
both the unitarity constraint and the experimental constraint are
the {\it available} anomalous coupling constants that an existing
heavy neutral Higgs boson can have.

Unlike what we did in the last section, we take account here the
Higgs decay rates and the Higgs width to full leading order in
perturbation, and we keep the nonvanishing weinberg angle, i.e.,
we use (\ref{csHZgamma}) and (\ref{csHZZ}) rather than
(\ref{HZgamma}) and (\ref{HZZ}) for ${\cal L}^{(6)}_{HZ\gamma}$
and ${\cal L}^{(6)}_{HZZ}$. In our numerical analysis, we take
FeynRules 2.0 \cite{FeynRules2.0} in our analysis code, and we use
MADGRAPH5 \cite{MADGRAPH5} to calculate the Higgs production and
decay rates.

In our effective couplings, there are altogether seven unknown
parameters, namely $C^{}_t, \rho^{}_h, \rho^{}_H, f^{}_W,
f^{}_{WW}, f^{}_B$ and $f^{}_{BB}$. So the analysis is rather
complicated. From Eq.\,(\ref{g(f)}), we see that $f^{}_B$ and
$f^{}_{BB}$ do not appear in the $HWW$ vertex, and they appear in
the $HZZ$ vertex with the suppression factors $s^2$ and $s^4$,
respectively. So their contributions to $VV$ scattering and $pp\to
VH^\ast\to VVV$ studied in our next paper are negligibly small.
They are mainly related to the decays $H\to\gamma\gamma$ and $H\to
Z\gamma$. However, for the heavy Higgs boson with $M^{}_H\ge 400$
GeV in our study, all the decay channels $H\to WW, H\to ZZ$ and
$H\to t\bar t$ are open, so that the two decay channels
$H\to\gamma\gamma$ and $H\to Z\gamma$ are relatively not so
important. Since we are not aiming at doing precision
calculations, we may take certain approximation to avoid dealing
with $f^{}_B$ and $f^{}_{BB}$ in the analysis to simplify it.

We then examine the ATLAS and CMS results of the strength
$\mu=\left.\sigma/\sigma_{SM}\right|^{}_{95\%~CL}$ for the decay
channels $H\to \gamma\gamma$
\cite{ATLAS_PRL12}\cite{CMS_PAS_HIG_13_001} and $H\to Z\gamma$
\cite{ATLAS_1402.3051}\cite{CMS_1307.5515}. Unfortunately, the
data only exist below 150 GeV which does not include the range
$M^{}_H\ge 400$ GeV in our study. So we can only make a
speculation of the situation in the range above 150 GeV. We see
from the results in
Refs.\,\cite{ATLAS_PRL12}\cite{CMS_PAS_HIG_13_001}\cite{ATLAS_1402.3051}\cite{CMS_1307.5515}
that the trend of the ATLAS and CMS results below 150 GeV is that
the experimental curves tend to gradually go closer to the $\mu=1$
axis. So we roughly estimate that they may keep this situation
above 150 GeV. This means that there is no evidence of needing
significant anomalous couplings in the $H\gamma\gamma$ and
$HZ\gamma$ couplings, i.e., we just neglect the anomalous coupling
constants of the effective $H\gamma\gamma$ and $HZ\gamma$
interactions. We frist see from Eq.\,(\ref{Hgammagamma}) that
neglecting the anomalous coupling constant in
Eq.\,(\ref{Hgammagamma}) means
\bea                                      %(32)
f^{}_{BB}\approx -f^{}_{WW}.
\label{f_BB-f_WW}                   %
\eea                                 %
We then see from Eq.\,(\ref{csHZgamma})
 that there are two terms in it. The first term
is proportional to $(q^{}_\mu q^{}_\nu-q^2g^{}_{\mu\nu})A^\mu$
which vanishes for on-shell photon. Thus neglecting the anomalous
coupling constant in Eq.\,(\ref{csHZgamma}) means
\bea                                      %(33)
f^{}_{B}\approx f^{}_{W}-4f^{}_{WW}.
\label{f_B-f_Wf_WW}                   %
\eea                                                                 %

Eqs.\,(\ref{f_BB-f_WW}) and (\ref{f_B-f_Wf_WW}) serve as two
constraints on $f^{}_{BB}$ and $f^{}_B$, expressing them in terms
of $f^{}_W$ and $f^{}_{WW}$. Then we have only five unknown
coupling constants left, namely $C^{}_t,\,\rho^{}_h,\,\rho^{}_H,\,
f^{}_W$ and $f^{}_{WW}$, as in the last section.

Next we look at ${\cal L}^{(6)}_{HWW}$ and ${\cal L}^{(6)}_{HZZ}$.
We see from (\ref{HWW}) that ${\cal L}^{(6)}_{HWW}$ does not
contain $f^{}_B$ and $f^{}_{BB}$, so it is unaffected by the
approximations (\ref{f_BB-f_WW}) and (\ref{f_B-f_Wf_WW}). However,
${\cal L}^{(6)}_{HZZ}$ does contain $f^{}_B$ and $f^{}_{BB}$. With
the approximations (\ref{f_BB-f_WW}) and (\ref{f_B-f_Wf_WW}),
Eq.\,(\ref{csHZZ}) becomes
\bea                           %(34)
&&\hspace{-0.4cm}{\cal L}^{(6)}_{HZZ}=
\frac{gM_W\rho^{}_H}{2c^2\Lambda^2}\bigg[\frac{1}{2}(f^{}_W-4s^2f^{}_{WW})\nonumber\\
&&\hspace{-0.2cm}\times (q_\mu q_\nu - q^2g_{\mu\nu}+k_\mu k_\nu -
k^2g_{\mu\nu})\nonumber\\
&&\hspace{-0.2cm}+(f^{}_W-2f^{}_{WW}) (q_\nu k_\mu -
g_{\mu\nu}q\cdot k ) \bigg] Z^\mu Z^\nu H.
\label{csHZZ'}                               %
\eea                                           %

Now we consider the CMS and ATLAS exclusion bounds on SM Higgs
boson \cite{CMS_HIG_13_002}\cite{ATLAS_CMS12}\cite{others}. The
strongest one is  CMS result obtained from the $H\to ZZ\to 4\ell$
channel \cite{CMS_HIG_13_002}. In this section, we mainly consider
this strongest bound, and we also take account of other weaker
bounds \cite{others} when considering the size of the {\it
available} range for $f^{}_W$ and $f^{}_{WW}$.

The strongest CMS exclusion bound is given in the Higgs mass range
up to 1 TeV. Its feature is that the experimental curve goes
rapidly away from the $\mu=1$ axis (below $\mu=1$) above 120 GeV,
and fluctuates in the range between 140 GeV and 400 GeV, and then
goes relatively smoother towards the $\mu=1$ axis up to 1 TeV. In
view of the significant fluctuations below 400 GeV, we shall take
$M^{}_H$=400 GeV, 500 GeV and 800 GeV as examples to do the
two-parameter analysis. The parameters in these examples are:\\\\
{\bf i, 400II}: $M^{}_H=400$ GeV, $C_t=0.5$ (Type-II),\\ \null~~~~
$\rho^{}_h=0.9$, $\rho^{}_H=0.4$,\\
\null\noindent {\bf ii, 500I}: $M^{}_H=500$ GeV, $C_t=1$ (Type-I),
$\rho^{}_h=0.9$,\\ \null~~~~ $\rho^{}_H=0.4$,\\
\null\noindent {\bf iii, 500II}: $M^{}_H=500$ GeV, $C_t=0.6$
(Type-II), $\rho^{}_h=0.8$,\\ \null~~~~ $\rho^{}_H=0.6$,\\
\null\noindent {\bf iv, 800I}: $M^{}_H=800$ GeV, $C_t=1$ (Type-I),
$\rho^{}_h=0.8$,\\ \null~~~~ $\rho^{}_H=0.6$,\\
\null\noindent {\bf v, 800II}: $M^{}_H=800$
GeV, $C_t=0.2$ (Type-II), $\rho^{}_h=0.9$,\\ \null~~~~ $\rho^{}_H=0.25$,\\

When calculating the strength $\mu$ for $H\to ZZ\to 4\ell$, we
need to calculate
\bea                                 %(35)
&&\sigma=\sigma(pp\to HX)B(H\to ZZ\to 4\ell),\nonumber\\
&&\hspace{-0.5cm} B(H\!\to \!ZZ\!\to \!4\ell)=\frac{\Gamma(H\!\to
\!ZZ\!\to \!4\ell)}{\Gamma(H\!\to\! ZZ)+\Gamma(H\!\to\!
WW)+\cdots}.~~~~      %
\label{BR}                        %
\eea                                %
The total decay width $\Gamma(H\!\to\! ZZ)+\Gamma(H\!\to\!
WW)+\cdots$ needs further discussion. Apart from the decay modes
related to the effective coupling mentioned in Sec.\,II, there can
also be the decay mode $H\to hh$ caused by an effective coupling
$\lambda v^{}_H Hhh$ (note that $H$ is the lightest heavy Higgs
boson so that it can not decay to other heavy Higgs bosons). For
$H$ with $M^{}_H\ge 400$ GeV, all the decay channels $H\to
WW,\,H\to ZZ$ and $H\to t\bar t$ are open. Since $M^{}_h$ is
larger than $M^{}_W$ and $M^{}_Z$, the phase space in $H\to hh$ is
smaller than those in $H\to WW$ and $H\to ZZ$. Thus the mode $H\to
hh$ does not play an important role in the total width. Since we
are not aiming at doing precision calculations, we can make the
approximation of neglecting the $H\to hh$ mode in the total decay
width of $H$ to avoid introducing a new unknown parameter
$\lambda$. In this approximation, our obtained total decay width
of $H$ is smaller than its actual value. This makes the obtained
exclusion constraint on $H$ stronger than what it actually is.
Thus our approximate calculation is a conservative calculation,
i.e., the required values of $f^{}_W$ and $f^{}_{WW}$ from our
approximate exclusion bound are more than enough for avoiding
being excluded by the actual exclusion bound. This guarantees that
a heavy Higgs boson $H$ with the obtained allowed values of
$f^{}_W$ and $f^{}_{WW}$ is definitely not excluded by the CMS
exclusion bound \cite{CMS_HIG_13_002}.

Now we present our two-parameter numerical analysis results.

\null\noindent{\it 1. $M^{}_H=$400 GeV}

As we have mentioned, the exclusion bound is very strong at
$M^{}_H=400$ GeV. Our numerical analysis shows that, for the case
of Type-I, a NS heavy neutral Higgs boson with $M^{}_H=400$ GeV
can hardly avoid being excluded. Of course, if we take $\rho^{}_H$
to be small enough, it may help. But a heavy neutral Higgs boson
with so small gauge interactions is not considered in this study,
and will be considered elsewhere.

\begin{figure}[h]                                 %Fig.3
\centering \bc
\hspace{0.5cm}\includegraphics[width=6.5cm,height=5cm]{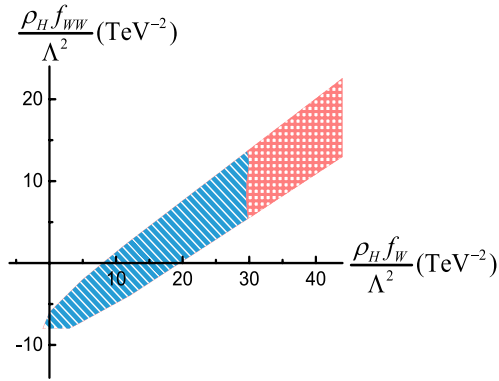}
 \ec
\vspace{-0.6cm} \caption{Obtained experimental bound on $f^{}_W$
and $f^{}_{WW}$ in the case of 400II. The blue shaded region is
the available region.}
\label{EB400II}                                %
\end{figure}

Now we consider case of 400II. The small $C_t$ reduces the Higgs
production cross section by gluon fusion $\sigma(pp\to HX)$, so
that the requirement of reducing $B(H\to ZZ\to 4\ell)$ is milder,
and it is possible to find out the available values of $f^{}_W$
and $f^{}_{WW}$. The result of our two-parameter numerical
analysis is shown in FIG.\,\ref{EB400II}. The shaded region means
the values of $f^{}_W$ and $f^{}_{WW}$ which can sufficiently
reduce the branching ratio $B(H\to ZZ\to 4\ell)$ such that the
heavy neutral Higgs boson is not excluded by the CMS exclusion
bound. Considering further the unitarity bound in
FIG.\,\ref{UB}(b), we find that {\it the real available region
(consistent with the unitarity bound) is the part shaded in blue}.

\null\noindent{\it 2. $M^{}_H$=500 GeV}

For $M^{}_H=500$ GeV, the SM Higgs exclusion bound is looser. We
take two sets of parameters as examples, namely 500I (Type-I) case
with $C_t=1,\, \rho^{}_h=0.9,\, \rho^{}_H=0.4$; and 500II
(Type-II) case with $C_t=0.6,\,\rho^{}_h=0.8,\,\rho^{}_H=0.6$.

\begin{figure}[h]                                 %Fig.4
\centering \bc
\hspace{0.5cm}\includegraphics[width=7cm,height=4cm]{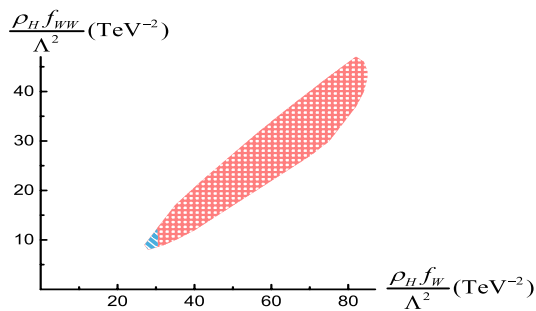}
 \ec
\vspace{-0.6cm} \caption{Obtained experimental bound on $f^{}_W$
and $f^{}_{WW}$ in the case of 500I. The blue shaded region is the
available region.}
\label{EB500I}                                %
\end{figure}

We first look at the 500I case. The result of our two-parameter
numerical analysis is shown in FIG.\,\ref{EB500I} in which the
shaded region is the region of $f^{}_W$ and $f^{}_{WW}$ making the
heavy neutral Higgs boson $H$ not excluded by the CMS exclusion
bound, and {\it the small part shaded in blue is consistent with
the unitarity bound shown in FIG.\,\ref{UB}(b), i.e., the
available region}. Note that this is also in the first quadrant of
the $f^{}_W$-$f^{}_{WW}$ plane.

\begin{figure}[h]                                 %Fig.5
\centering \bc
\hspace{0.5cm}\includegraphics[width=7cm,height=5cm]{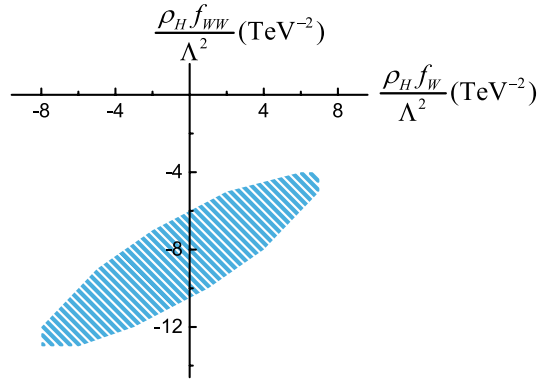}
 \ec
\vspace{-0.6cm} \caption{Obtained experimental bound on $f^{}_W$
and $f^{}_{WW}$ in the case of 500II. The blue shaded region is
the available region.}
\label{EB500II}                                %
\end{figure}

Next we look at the 500II case. The result of our two-parameter
numerical analysis is shown in FIG.\,\ref{EB500II} in which {\it
the blue shade region is the available region (the whole region is
consistent with the unitarity bound shown in FIG.\,\ref{UB}(a))}.
These available region is in the third and fourth quadrants. Since
the value $\rho^{}_H=0.6$ is larger than that in the 500I case,
the needed values of $f^{}_W$ and $f^{}_{WW}$ for sufficiently
reducing $B(H\to ZZ\to 4\ell)$ in the first quadrant are so large
that they exceed the unitarity bound shown in FIG.\,\ref{UB}(a).
Thus only the region shown in FIG.\,\ref{EB500II} is really
available.

\null\noindent{\it 3. $M^{}_H$=800 GeV}

We see from the CMS exclusion bound \cite{CMS_HIG_13_002} that the
exclusion bound at $M^{}_H$=800 GeV is very loose, so that almost
all values of $f^{}_W$ and $f^{}_{WW}$ are available to make the
heavy neutral Higgs boson not excluded by the CMS exclusion bound.
In the 800I case, $C_t=1$, the total decay width of the 800 GeV
heavy neutral Higgs boson is quite large that it is not possible
to see a resonance bump at the LHC, but it is still possible to
detect it by measuring the cross section. In the 800II case, a
sufficiently small value of $C_t$ will make the total decay width
small enough that a resonance bump can be seen at the LHC.

To understand why the available regions in FIGs.\,\ref{EB400II},
\ref{EB500I} and \ref{EB500II} are so different, let us look at
how $f^{}_W$ and $f^{}_{WW}$ affect $\Gamma(H\to WW)$ and
$\Gamma(H\to ZZ)$. Below are our obtained results of $\Gamma(H\to
WW)$ and $\Gamma(H\to ZZ)$.
\bea                            %(36)
&&\Gamma(H\to WW)\approx
\frac{g^2\rho_H^2M_H^3}{64\pi M_W^2}\left[\left(1-\frac{M_W^2}{\Lambda^2}f^{}_W\right)^2\right.\nonumber\\
&&~~\left.+2\frac{M_W^4}{\Lambda^4}(f^{}_W-2f^{}_{WW})^2+O\left(\frac{M_W^2}{M_H^2}\right)\right],
\label{GammaWW}                  %
\eea
\bea                            %(37)
&&\hspace{-0.4cm}\Gamma(H\to ZZ)\approx
\frac{g^2\rho_H^2M_H^3}{128\pi M_W^2}\left[\left(1\!-\!\frac{M_Z^2}{\Lambda^2}(f^{}_W\!-\!4s^2f^{}_{WW})\right)^2\right.\nonumber\\
&&~~\left.+2\frac{M_Z^4}{\Lambda^4}(f^{}_W-2f^{}_{WW})^2+O\left(\frac{M_W^2}{M_H^2}\right)\right],
\label{GammaZZ}                  %
\eea                             %

First of all, we see from (\ref{GammaWW}) and (\ref{GammaZZ})
that, if $f^{}_W$ and $f^{}_{WW}$ are in the second quadrant of
the $f^{}_W$-$f^{}_{WW}$ plane, i.e., $f^{}_W<0,\,f^{}_{WW}>0$,
they always increase $\Gamma(H\to WW)$ and $\Gamma(H\to ZZ)$, and
$\Gamma(H\to ZZ)$ is increased more than $\Gamma(H\to WW)$ is. In
this case, $B(H\to ZZ\to 4\ell)$ is always increased, so that the
heavy Higgs boson $H$ is definitely excluded by the CMS exclusion
bound, i.e., {\it there is no available region of $f^{}_W$ and
$f^{}_{WW}$ in the second quadrant of the $f^{}_W$-$f^{}_{WW}$
plane}. It is so in  FIGs.\,\ref{EB400II}, \ref{EB500I} and
\ref{EB500II}.

Next we look at the case that
$\left|f^{}_W\right|,\,\left|f^{}_{WW}\right|<\Lambda^2/M_W^2$
with $f^{}_W-2f^{}_{WW}$ not too large. We see from
(\ref{GammaWW}) and (\ref{GammaZZ}) that, for $f^{}_W$ and
$f^{}_{WW}$ in the first quadrant ($f^{}_W>0,\,f^{}_{WW}>0$),
$\Gamma(H\to WW)$ and $\Gamma(H\to ZZ)$ are all decreased, and
$\Gamma(H\to WW)$ is decreased more than $\Gamma(H\to ZZ)$ is. So
that $B(H\to ZZ\to 4\ell)$ is increased, i.e., {\it there is no
available region of $f^{}_W$ and $f^{}_{WW}$ in the first quadrant
of the $f^{}_W$-$f^{}_{WW}$ plane}. However, in the third quadrant
($f^{}_W<0,\,f^{}_{WW}<0$) and the fourth quadrant
($f^{}_W>0,\,f^{}_{WW}<0$), either $\Gamma(H\to WW)$ is increased
more than $\Gamma(H\to ZZ)$ is, or $\Gamma(H\to WW)$ is decreased
less than $\Gamma(H\to ZZ)$ does. Thus in these two quadrants,
$B(H\to ZZ\to 4\ell)$ is reduced, so that {\it there can be
available region of $f^{}_W$ and $f^{}_{WW}$ in the third and
fourth quadrants of the $f^{}_W$-$f^{}_{WW}$ plane}. This is just
the situation in
FIG.\,\ref{EB500II}. In the special case of 400II%$M^{}_H=400$ GeV
with $C_t=0.5$ which significantly reduces the Higgs production
cross section $\sigma(pp\to HX)$, in addition to the third and
fourth quadrants, there can also be available region in the first
quadrant even $B(H\to ZZ\to 4\ell)$ is increased a little there.
Thus in this special case, {\it there can be available regions in
the first, third, and fourth quadrants}. This is the situation in
FIG.\,\ref{EB400II}.

 We then look at the case that
$\left|f^{}_W\right|,\,\left|f^{}_{WW}\right|\sim\Lambda^2/M_W^2$.
In this case, we should examine both the first and second terms in
(\ref{GammaWW}) and (\ref{GammaZZ}). In the first quadrant, the
first terms are quite small, and the second terms (proportional to
$f^{}_W-2f^{}_{WW}$) can also be small when $f^{}_W\approx
2f^{}_{WW}$, while the total decay rate [the denominator in
Eq.\,(\ref{BR})] is not reduced so much since $\Gamma(H\to t\bar
t)$ is not so small. So, in this case, $B(H\to ZZ\to 4\ell)$ can
be sufficiently reduced. In the fourth quadrant, the second terms
are not small enough, and in the third quadrant, the first terms
are not small enough. So that in the third and fourth quadrants
$B(H\to ZZ\to 4\ell)$ cannot be sufficiently reduced. Thus in this
case {\it there can be available region of $f^{}_W$ and
$f^{}_{WW}$ only in the first quadrant of the $f^{}_W$-$f^{}_{WW}$
plane}. This is the situation in FIG.\,\ref{EB500I}.

When $\left|f^{}_W\right|$ and $\left|f^{}_{WW}\right|$ become
larger and larger, the constant terms (independent of $f^{}_W$ and
$f^{}_{WW}$) in (\ref{GammaWW}) and (\ref{GammaZZ}) are less and
less important. In this case, $\Gamma(H\to WW)$ and $\Gamma(H\to
ZZ)$ all increase, and they are different only by the term
containing $4s^2f^{}_{WW}$. it can be shown that, in this case,
\bea                        %(38)
\Gamma(H\to ZZ)\nless 0.2\,\Gamma(H\to WW),
\label{ZZ/WW}                %
\eea                        %
or
\bea                            %(39)
\frac{\Gamma(H\to ZZ)}{\Gamma(H\to WW)+\Gamma(H\to ZZ)}\nless
0.17.
\label{branching}                       %
\eea                              %
Comparing the corresponding SM values, our detailed analysis shows
that, for $M^{}_H$= 400 GeV and 500 GeV, this is not small enough
for sufficiently reducing $B(H\to ZZ\to 4\ell)$ to avoid being
excluded by the CMS bound in Ref.\,\cite{CMS_HIG_13_002}. Thus the
available values of $\left|f^{}_W\right|$ and
$\left|f^{}_{WW}\right|$ cannot be arbitrarily large. This is why
the available regions in FIG.\,\ref{EB400II}, FIG.\,\ref{EB500I},
and FIG.\,\ref{EB500II} are all closed regions.

Finally, we would like to add a discussion on whether it is
reasonable to simply apply the CMS exclusion bound to our examples
with new physics interactions as what we did above. We know that
the detection efficiency of the detector depends on specific
interactions, and the detection efficiency of the CMS exclusion
bound in Ref.\,\cite{CMS_HIG_13_002} is for the SM interaction. We
shall take 400II with $\rho^{}_Hf^{}_W/\Lambda^2=30$ TeV$^{-2}$
and $\rho^{}_Hf^{}_{WW}/\Lambda^2=10$ TeV$^{-2}$., 500I with
$\rho^{}_Hf^{}_W/\Lambda^2=30$ TeV$^{-2}$,
 and $\rho^{}_Hf^{}_{WW}/\Lambda^2=10$ TeV$^{-2}$, and 500II with
 $\rho^{}_Hf^{}_W/\Lambda^2=6$ TeV$^{-2}$ and $\rho^{}_Hf^{}_{WW}/\Lambda^2=-5$ TeV$^{-2}$
 as
examples to calculate how much their detection efficiencies
deviate from the that with the SM interaction.

 We shall make a calculation to study how much such deviations actually are in
detecting $pp\to H\to ZZ\to \ell^+\ell^-\ell^+\ell^-$ at the 8 TeV
LHC. We use DELPHES 3 \cite{DELPHES3} to roughly simulate the
detector. We use MADGRAPH 5 to do the simulation, and use
MadAnalysis to obtain the efficiency.

In our calculation, we have chosen 60 GeV $<M(\ell^+\ell^-)<$120
GeV to guarantee that the two final states $\ell^+$ and $\ell^-$
are from the decay of a $Z$ boson. We have also chosen 200
GeV$<M(\ell^+\ell^-\ell^+\ell^-)<$ 600 GeV and 300
GeV$<M(\ell^+\ell^-\ell^+\ell^-)<$ 700 GeV to guarantee the final
state $\ell^+\ell^-\ell^+\ell^-$ are from the decay of our heavy
Higgs bosons under consideration.

The obtained detection efficiency for detecting $H\to ZZ\to
\ell^+\ell^-\ell^+\ell^-$ is listed in
TABLE~\ref{detection-efficiency}.

\begin{widetext}

\begin{table}[h]                      %Tab1
\tabcolsep 15pt             %
\label{detection-efficiency}     %
\caption{Comparison of the detection efficiencies between the SM
and our examples: 400II with $\rho^{}_Hf^{}_W/\Lambda^2=30$
TeV$^{-2}$ and $\rho^{}_Hf^{}_{WW}/\Lambda^2=10$ TeV$^{-2}$., 500I
with $\rho^{}_Hf^{}_W/\Lambda^2=30$ TeV$^{-2}$,
 and $\rho^{}_Hf^{}_{WW}/\Lambda^2=10$ TeV$^{-2}$, and 500II with
 $\rho^{}_Hf^{}_W/\Lambda^2=6$ TeV$^{-2}$ and $\rho^{}_Hf^{}_{WW}/\Lambda^2=-5$ TeV$^{-2}$.}
\begin{tabular}{cccccc}
\hline\hline\\
 & 400II & SM ($M^{}_H=400$ GeV) & 500I &500II &SM ($M^{}_H=500$ GeV) \\
\hline\\
detection efficiency& $17.9\% $ & $ 17.7\% $ & $18.6\% $ & $18.8\% $ & $ 17.6\% $\\
\hline\hline
\end{tabular}
\label{detection-efficiency}           %
\end{table}
\end{widetext}

We see that, for 400II, the new interaction causes a relative
change of the efficiency with respect to the SM efficiency by
$(17.9\%-17.6\%)/17.6\%= 2\%$. For 500I and 500II, the
corresponding relative changes of the efficiency are
$(18.6\%-17.6\%)/17.6\%= 6\%$ and $(18.8\%-17.6\%)/17.6\%= 7\%$,
respectively. Since we are not aiming at doing precision
calculations, a few percent change will not affect our main
conclusions in simply applying the CMS exclusion bound to our
examples..

\section{General features of studying the LHC signatures of $\bm H$}

For the study of the LHC signatures of $H$ at the 14 TeV LHC, we
do not suggest taking the conventional on-shell Higgs production,
used in studying the properties of the 125--126 GeV Higgs boson,
to probe the anomalous heavy Higgs boson, The reason is the
following. Comparing Eq.\,(\ref{LHeff}) with
Eq.\,(\ref{dim-4Hcoupling}) in Sec.\,II, we see that the dim-6
interaction contains an extra factor $k^2/\Lambda^2$ relative to
the dim-4 interaction, coming from the extra derivatives in
Eq.\,(\ref{LHeff}). Here $k$ is a typical momentum of the order of
the momentum of the Higgs boson. In on-shell Higgs production,
$k^2\sim M_H^2$. Taking $M^{}_H=500$ GeV as an example,
$k^2/\Lambda^2\sim M_H^2/\Lambda^2=0.25/9=0.03$. Thus the
contribution of the dim-6 interaction is only a very tiny portion
of the total contribution, so that it is hard to detect the dim-6
interaction effect in on-shell Higgs production.

Instead, in this paper, we suggest taking $VV$ scattering and
$pp\to VH^\ast\to VVV$ as sensitive processes for probing the
anomalous heavy Higgs boson at the 14 TeV LHC. These processes
contain off-shell heavy Higgs contributions. In the tail with
energy higher than the resonance peak, $k^2/\Lambda^2$ can be
larger. Although the tail with much higher energy than the
resonance is seriously suppressed by the parton distribution
(e.g., the region $k^2\alt \Lambda^2$ is almost completely
suppressed), the remaining high energy tail can still enhance the
contribution of the dim-6 interaction as we shall see in
Secs.\,VI--VIII. Furthermore, each of these two processes contains
two $HVV$ vertices. This makes the cross sections more sensitive
to the anomalous couplings than in on-shell Higgs production.

Although the two suggested processes are weak-interaction
processes with not so large cross sections, the signal to
background ratio can be effectively improved by imposing a series
of proper cuts. So that the integrated luminosity needed for
$3\sigma$ and $5\sigma$ deviations are not so high (e.g., see
TABLE\,\ref{VH-IL}) in Sec.\,VII.

That weak-boson scattering can be a sensitive process for
detecting an anomalous Higgs boson at the LHC was first pointed
out in Ref.\,\cite{ZKHY03}, in which the effective couplings for a
single-Higgs system, and the pure leptonic decay mode of the final
state $W$ bosons was considered. It showed that the required
integrated luminosity was high. Ref.\,\cite{QKLZ09} studied the
same problem but with the semileptonic decay channel (one of the
final state $W$ boson decays to leptons and the other $W$ boson
decays to jets), and showed that the required integrated
luminosity was significantly reduced.

Thus we shall study the semileptonic decay mode in both the
weak-boson scattering and $pp\to VH^\ast\to VVV$ processes, i.e.,
$pp\to V V j^f_1j^f_2\to \ell^+ \nu^{}_\ell j_1j_2j^f_1j^f_2$
($j^f_1,\,j^f_2$ stand for forward jets) in weak-boson scattering,
and $pp\to VH^\ast\to VVW\to \ell^+\nu^{}_\ell
j^{}_1j^{}_2j^{}_3j^{}_4$ in the $pp\to VH^\ast\to VVV$ process.
Since there are several jets in the final states, parton-level
calculation is not adequate. We shall do the calculation to the
hadron level.

We take the CTEQ6.1 parton distribution functions \cite{CTEQ6.1},
and use MADGRAPH5 \cite{MADGRAPH5} to do the full tree-level
simulation. The parton shower and hadronization are calculated
with PYTHIA6.4 \cite{PYTHIA6.4}, and the anti-$k^{}_T$ algorithm
\cite{antik_T} in DELPHES 3 \cite{DELPHES3} is used for the
formation of jets with $R=0.7$ \cite{CMS2011}. We also use DELPHES
3 to simulate the detecting efficiency of the detector. We take
the five examples in Sec.\,IV to do the simulation, and take the
acceptance of the detector listed in TABLE\,\ref{acceptance}.

\begin{table}[h]                      %Tab2
\tabcolsep 20pt
%\begin{center}
\caption{The detector acceptance.}
\begin{tabular}{cccc}
\hline \hline
        & $|\eta|^{}_{max}$ & $P^{}_{T {\rm min}}$ \\
\hline
$\mu$   & 2.4            & 10GeV      \\
$e$     & 2.5            & 10GeV      \\
jet     & 5              & 20GeV      \\
photon  & 2.5            & 0.5GeV     \\
\hline \hline
\end{tabular}
%\caption{The detector acceptance.}
\label{acceptance}                           %
%\end{center}
\end{table}

%\null\vspace{-0.5cm}
In each process, we regard the contributions by the heavy Higgs
boson $H$ as the signal, other contributions without $H$ as
backgrounds. Among the backgrounds processes, the process with the
same initial- and final-state is regarded as the irreducible
background (IB), others are reducible backgrounds (RB). The signal
and the IB should be calculated together since they have
interference. Let $\sigma$ be the total cross section. The
background and the signal cross sections are then defined as
\bea                                     %(40)
&&
\sigma^{}_B=\sigma(C_t=1,\rho^{}_h=1,\rho^{}_H=0,f^{}_W=0,f^{}_{WW}=0),\nonumber\\
&&\sigma^{}_S=\sigma-\sigma^{}_B.
\label{sigma}                             %
\eea                                     %
For an integrated luminosity ${\sf L}^{}_{int}$, The signal and
background event numbers are $N^{}_S={\sf
L}^{}_{int}\sigma^{}_S,\,\,N^{}_B={\sf L}^{}_{int}\sigma^{}_B$. In
this paper, we take the Poisson distribution approach to determine
the statistical significance $\sigma^{}_{stat}$. The general
Poisson probability distribution reads
\bea                           %(41)
\displaystyle
P^{}_B&=&\sum_N\displaystyle e^{-N^{}_B}\frac{N_B^N}{N!},\nonumber\\
&&N=N^{}_S+N^{}_B,N^{}_S+N^{}_B+1,\cdots,\infty.
\label{P_B}                                    %
\eea                                          %
Comparing the obtained value of $1-P^{}_B$ with the probability of
the signal in the Gaussian distribution, we can find out the
corresponding value of $\sigma^{}_{stat}$ \cite{PDG}. The value of
$\sigma^{}_{stat}$ obtained in this way approaches to the simple
form
\bea                              %(42)
\sigma^{}_{stat}=\frac{N^{}_S}{\sqrt N^{}_B}
\label{stat}                           %
\eea                                     %
 when $N^{}_S$ and $N^{}_B$
are sufficiently large.

\section{Probing Heavy Neutral Higgs Bosons via Weak-Boson
scattering}

\begin{widetext}

\begin{figure}[h]                                 %Fig.6
\centering   \bc
\includegraphics[width=16cm,height=8.5cm]{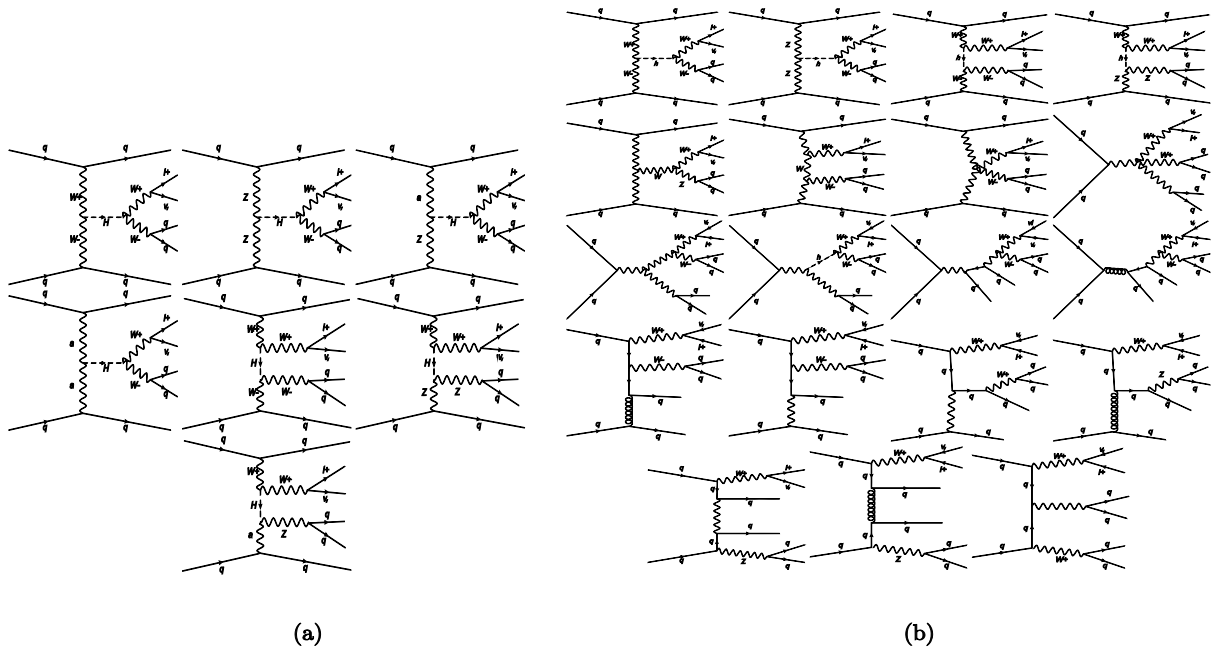}
 \ec \vspace{-0.4cm} \caption{Feynman diagrams in weak-boson scattering. (a) the
 signal, (b) examples of the IB.}
 \label{VV-sig&IB}                       %
\end{figure}
\end{widetext}

In this section, we study  the semileptonic mode of weak-boson
scattering, $pp\to V V j^f_1j^f_2\to \ell^+\nu^{}_\ell j_1j_2
j^f_1j^f_2$. We first look at the Feynman diagrams of the signal,
IB, and RBs in this process. Feynman diagrams for the signal and
examples of the IB are shown in FIG.\,\ref{VV-sig&IB}

These two kinds of diagrams in FIG.\,\ref{VV-sig&IB}(a) and
FIG.\,\ref{VV-sig&IB}(b) should be calculated together since they
have interference.

\begin{figure}[h]                                 %Fig.7
\centering   \bc
\includegraphics[width=7cm,height=6.5cm]{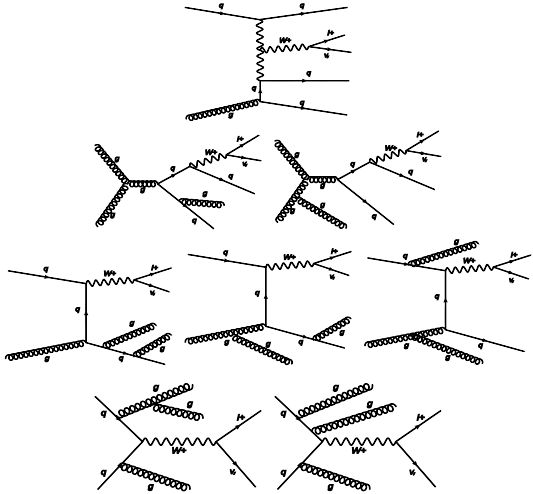}
 \ec \vspace{-0.4cm} \caption{Feynman diagrams for QCD backgrounds %. (a)
of $W+3j$.}%, (b) $WV+$2-jets.}
 \label{QCDRBW3j}                       %
\end{figure}

\begin{figure}[h]                                 %Fig.8
\centering   \bc
\includegraphics[width=5.5cm,height=4cm]{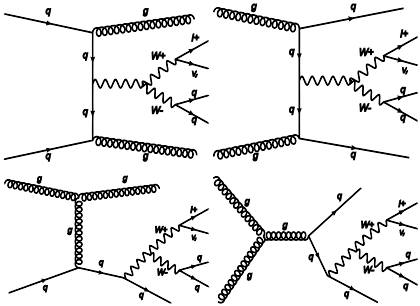}\\
 \ec \vspace{-0.4cm} \caption{Feynman diagrams for QCD backgrounds
of $WV+2j$.}
 \label{QCDRBWV2j}                       %
\end{figure}

Apart from the IB, there are two kinds of RBs, namely the
so-called QCD backgrounds and top-quark backgrounds
\cite{Bagger9495}. Note that the two jets $j^{}_1$ and $j^{}_2$
from $W$ decay mainly behave as a ``single'' energetic fat jet $J$
along the $W$ direction \cite{Butterworth}\cite{CMS_JME_13_006}
since the final state $W$ is very energetic. This is the reason
why we take $R=0.7$ in the anti-$k_T$ algorithm. In this case, the
important QCD backgrounds which can mimic the signal at the hadron
level are the the inclusive $W+3j$ (with $W\to \ell^+\nu^{}_\ell$,
and the three jets mimic the fat jet $J$ and the two forward jets)
and the $WV+2j$ (with $W\to \ell^+\nu^{}_\ell$, $V\to J$, and the
two jets mimic the two forward jets). The parton-level Feynman
diagrams of these two QCD backgrounds are shown in
FIGs.\,\ref{QCDRBW3j} and FIGs.\,\ref{QCDRBWV2j}. These two QCD
backgrounds have been discussed in Ref.\,\cite{QKLZ09}. In our
calculation, we match the partons with jets using the method in
Refs.\,\cite{MMP02}\cite{Aetal08} to obtain the inclusive $W+3j$
and inclusive $WV+2j$ backgrounds.

\begin{figure}[h]                                 %Fig.9
\centering   \bc
\includegraphics[width=5.5cm,height=4cm]{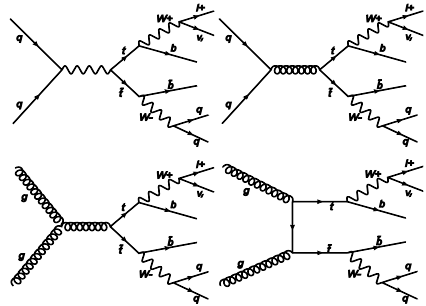}\\
 \ec \vspace{-0.4cm} \caption{Feynman diagrams for the top-quark backgrounds.}
\label{topB}                       %
\end{figure}

The top-quark background is $pp\to t\bar t\to W^+bW^-\bar{b}\to
\ell^+\nu^{}_\ell j^{}_1j^{}_2 b\bar{b}$ with $j^{}_1j^{}_2
b\bar{b}$ mimicking the two jets in $W$ decay and the two forward
jets. The Feynman diagrams of the top-quark background are shown
in FIB.\,\ref{topB}.

We shall take the following kinematic cuts, reflecting the
properties of the signal, to suppress the backgrounds and keep the
signal as much as possible.

\null\noindent{\bf cut1}: {\it Requiring an isolated lepton
$\ell^+\,(\mu^+,\,e^+)$ in the central rapidity region}
\bea                                  %(43)
&&N(\ell^+)=1,~~~~N(\ell^-)=0\nonumber\\
{\rm with}~~~~~~~~~~~~~~&&\left|\eta^{}_{\ell^+}\right|<2.
\label{leptoncut}                                 %
\eea                                             %
Since the signal lepton has larger probability to be in the
central rapidity region than the RBs do, this cut will suppress
the RBs relative to the signal. Furthermore, there can be fake
leptons ($\ell^+\,\,{\rm or}\,\,\ell^-$) coming from the decays of
the hadrons $\pi,\,\eta,\,J/\psi$, etc. in the hadronized jets.
This cut can also suppress the fake leptons.

\null\noindent{\bf cut2}: {\it $p^{}_T(leptons)$-cut}

Let ${\bm p}^{}_T(\ell^+)$ and~ $/\!\!\!\!{\bm p}^{}_T\equiv{\bm
p}^{}_T(\nu^{}_\ell)$ be the transverse momentum vectors of
$\ell^+$ and $\nu^{}_\ell$, respectively.
\begin{figure}[h]                                 %Fig.10
\centering   \bc
\includegraphics[width=5cm,height=3cm]{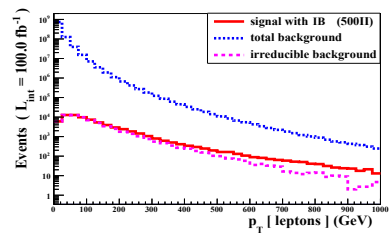}\\
 \ec \vspace{-0.4cm} \caption{$p^{}_T(leptons)$ distributions of signal+IB (red-solid),
 IB (pink-dotted) and total RBs (blue-small-dotted) in weak-boson scattering for the example 500II with ${\sf L}^{}_{int}=$100 fb$^{-1}$.}
 \label{p_T(leptons)}                       %
\end{figure}
Our simulation shows that a cut on
$p^{}_T(leptons)\equiv\left|{\bm p}^{}_T(\ell^+)+/\!\!\!\!{\bm
p}^{}_T\right|$ can effectively suppress both the IB and the RBs.
FIG.\,\ref{p_T(leptons)} plots the inclusive $p^{}_T(leptons)$
distributions of the signal plus IB (red-solid), the IB
(pink-dotted), and the total RBs (blue-small-dotted) for example
500II with ${\sf L}^{}_{int}=$100 fb$^{-1}$. We see from
FIG.\,\ref{p_T(leptons)} that taking a cut
\bea                                  %(44)
p^{}_T(leptons)>150\,{\rm GeV}
\label{p_T(l)cut}                                 %
\eea                                       %
can suppress both the IB and the total RBs, while keep the signal
as much as possible. It can also suppress fake leptons very
effectively since the scale of the transverse momenta of fake
leptons is of the order of the hadronization scale which is much
smaller than the required $p^{}_T(leptons)$ in (\ref{p_T(l)cut}).

\null\noindent{\bf cut3}: {\it Forward-jet cuts}

The signal has two clear forward jets $j^f_1$ and $j^f_2$ which
characterize the weak-boson fusion process, while in some RBs, the
jets which mimic $j^f_1$ and $j^f_2$ may not be forward. So that
we can set cuts reflecting the properties of $j^f_1$ and $j^f_2$
to suppress the RBs. There have been several ways of setting the
forward-jet cuts. We follow the way in Ref.\,\cite{Butterworth}
but with a little modification
\bea                                %(45)
&& p^{}_T(j^f)> 35~{\rm GeV},\nonumber\\
&&E(j^f)>300~{\rm GeV},\nonumber\\
&&2.0<|\eta(j^f)|<5,~~~~~~\eta(j^f_1)\eta(j^f_2)<0.
\label{forward jet cuts}                  %
\eea                                      %
In the cut for $|\eta(j^f)|$, we have taken account of the
acceptance of the detector (cf. TABLE\,\ref{acceptance}). Here,
instead of taking $p^{}_T(j^f)> 20~{\rm GeV}$ as in
Ref.\,\cite{Butterworth}, we take $p^{}_T(j^f)> 35~{\rm GeV}$ for
avoiding the pileup events.

In our simulation, we take the jet with most positive $\eta$ and
the jet with most negative $\eta$ to satisfy the rapidity
requirement in (\ref{forward jet cuts}).

\null\noindent{\bf cut4}: {\it Fat jet cuts}

In the signal, the fat jet $J$ (the jet with largest transverse
momentum) is the decay product of a $W$ boson, so that the
invariant mass $M(J)$ of $J$ should equal to $M^{}_W$. Considering
the resolution of the detector, we set the requirement
\bea                              %(46)
70\,{\rm GeV}<M(J)<100\,{\rm GeV}.
\label{M(J)}                         %
\eea                                 %
This requirement can effectively suppress the largest reducible
background $W+3j$ since, in $W+3j$, the largest $p^{}_T$ ordinary
jet $\hat j$ which mimics $J$ comes from the clustering of the
parton showers from a massless parton. For most of the
probability, its mass $M(\hat j)$ is much smaller than the
requirement (\ref{M(J)}).

Furthermore, in the signal, the fat jet $J$ and the isolate lepton
$\ell^+$ are decay products of the two $W$ bosons in $H$ decay.
With the cut (\ref{leptoncut}), we also set
\bea                              %(47)
|\eta^{}_J|<2
\label{eta(J)}                         %
\eea                                  %
to suppress the backgrounds.

\null\noindent{\bf cut5}: {\it Top-quark veto}

We see from FIG.\,\ref{topB} that, in a top-quark background
event, $t\to W^+ b\to \ell^+{\bar \nu}_\ell b$, $\bar t\to W^-
{\bar b}\to J{\bar b}$. So that, to identify a top-quark
background event, we can construct the invariant mass $M(J,\bar
b)$ to reconstruct the top quark. Experimentally, $M(J,\bar b)$
must be in the top-quark resonance region around $m^{}_t$. On the
other hand, if we construct $M(J,b)$ it will not be in the
top-quark resonance region. However, in the experiment, we can
just see three jets $J,\,j^{}_1,\,j^{}_2$ in the final state, and
cannot identify which one of $j^{}_1$ and $j^{}_2$ is the $\bar b$
jet. So we should construct two invariant masses  $M(J,j^{}_1)$
and $M(J,j^{}_2)$ to see if one of them is in the top-quark
resonance region to identify whether an event is a top-quark
background event. In FIG.\,\ref{VV-M(J,j)} we plot the
$M(J,j^{}_1)$ [or $M(J,j^{}_2)$] distribution from our simulation
including the signal plus IB (red-solid) and the top-quark
background (blue-dotted) distributions for the example 500II with
${\sf L}^{}_{int}=$100 fb$^{-1}$.
\begin{figure}[h]                                 %Fig.11
\centering   \bc
\includegraphics[width=6cm,height=3cm]{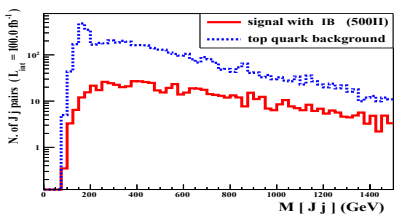}\\
 \ec \vspace{-0.4cm} \caption{$M(J,j)$ distributions of signal+IB (red-solid)
 and the top-quark background (blue-dotted) in weak-boson scattering for the example 500II with ${\sf L}^{}_{int}=$100 fb$^{-1}$.}
 \label{VV-M(J,j)}                       %
\end{figure}
We see that the top-quark resonance region is between 130 GeV and
240 GeV \cite{Butterworth}. So if, in an event, one of the
invariant masses $M(J,j^{}_1)$ and $M(J,j^{}_2)$ is in the region
\bea                               %(48)
{\rm 130\,GeV}<M(J,j)<{\rm 240\,GeV},
\label{topveto}                           %
\eea                                       %
we should veto the event. Equivalently, we only take the events in
which both $M(J,j^{}_1)$ and $M(J,j^{}_2)$ are outside the region
(\ref{topveto}). In this way, we can effectively veto the
top-quark background events.

Actually, there are more untagged jets apart from the tagged jets
$J$, $j^{}_1$ and $j^{}_2$ in the result of the anti-$k_T$
algorithm. For safety, we have also checked the constraint
(\ref{topveto}) for invariant masses of $J$ with all other
untagged jets.

\null\vspace{0.5cm} To see the efficiency of each cut, we list the
values of the cross sections [in fb] for signal plus IB (for the
five examples mentioned in Sec.\,I) and various backgrounds after
each cut in TABLE\,\ref{VV-efficiency}. We see that, with all
these cuts, the backgrounds can be effectively suppressed.

\begin{widetext}

\begin{center}
\begin{table}[h]                         %Tab3
\caption{Cut efficiencies expressed in terms of the tree-level
cross sections $\sigma^{}_{S+IB}$ and $\sigma^{}_B$ (in unit of
fb) in the weak-boson scattering process. The first five columns
are values of $\sigma^{}_{S+IB}$ for the five examples, and the
last four columns are values $\sigma^{}_B$ for four kinds of
backgrounds.}
\tabcolsep 9.6pt                        %
\begin{tabular}{ccccccccccc}
\hline \hline &&&$\sigma^{}_{S+IB}$&&&&&&$\sigma^{}_B$&\\
        & 400\uppercase\expandafter{\romannumeral2}  & 500\uppercase\expandafter{\romannumeral1} & 500\uppercase\expandafter{\romannumeral2}
        & 800\uppercase\expandafter{\romannumeral1} & 800\uppercase\expandafter{\romannumeral2} && IB &  W+jets & $t\bar{t}$ & WV+jets \\
\hline
without cuts &2085 &2037 &2009&1917&1996&&1925 &31500000&92000&7600 \\
{\bf cut1}   &759  &740  &726 &679 &705 &&669  &9360000 &35792&2506 \\
{\bf cut2}   &210  &209  &185 &149 &162 &&138  &44270   &5298 &499  \\
{\bf cut3}  &11.5 &11.0 &14.6&10.6&11.3 &&8.51 &370     &123  &13.7 \\
{\bf cut4}  &1.20 &1.28 &2.33&1.59&1.92 &&0.682 &5.47    &10.3 &1.53 \\
{\bf cut5}  &0.936&0.921&1.80&1.22&1.56 &&0.474 &3.49    &2.04
&0.81\\
%{\bf cut6}  &X&X&X&--&--&&X&X&X&X\\
 \hline \hline
\end{tabular}
\label{VV-efficiency}
\end{table}
\end{center}
\end{widetext}
We see that, before imposing the cuts, the $W+jets$ background is
larger than the signal plus IB by a factor of $1.5\times 10^4$.
After {\bf cut1}--{\bf cut5}, it is reduced to the same order of
magnitude as the signal plus IB.

Now the cross sections are of the order of 1 fb, so that for an
integrated luminosity of 50--100 fb$^{-1}_{}$, there can be
several tens to hundreds events which are detectable in the first
few years run of the 14 TeV LHC.

\begin{center}
\begin{table}[h]                   %Tab4                  %
\caption{Required integrated luminosity ${\sf L}^{}_{int}$ (in
unit of fb$^{-1}$) for the statistical significance of
$1\sigma,\,3\sigma$ and $5\sigma$ for the five examples in
weak-boson scattering.}
\tabcolsep 9pt                        %
\begin{tabular}{cccccc}
\hline \hline
 & & &${\sf L}_{int}$\,[fb$^{-1}_{}]$ & &\\
 &400II&500I&500II&800I&800II\\
 \hline
$1\sigma$&32 &34 &3.9 &12 &5.7 \\
$3\sigma$&288 &397 &35 &110 &52 \\
$5\sigma$&800 &852 &96 &306 &143 \\
 \hline \hline
\end{tabular}
\label{VV-IL}
\end{table}
\end{center}

From Eqs.\,(\ref{sigma})--(\ref{stat}), we obtain the following
required integrated luminosity for the statistical significance of
$1\sigma$, $3\sigma$ and $5\sigma$ for the five examples mentioned
in Sec.\,I (cf. TABLE\,\ref{VV-IL}).

 We see that examples 500II
and 800II are hopeful to be discovered (at the $5\sigma$ level) in
the first few years run of the 14 TeV LHC; while 800I can be
discovered (at the $5\sigma$ level), and 400I and 500I can have
evidences (at the $3\sigma$ level) for an integrated luminosity of
300 fb $^{-1}$ at the 14 TeV LHC.

Of course we have only taken account of the statistical error
here, and we leave the study of the systematic errors to the
experimentalists.

There is a missing neutrino in the final state. we take the method
of determining the neutrino longitudinal momentum from the
requirement of reconstructing the correct value of the $W$ boson
mass suggested by Ref.\,\cite{Butterworth}. There are two
solutions of the longitudinal momentum of the neutrino. we take
the solution with smaller $p^{}_z$ as is conventionally used
\cite{cms2013search}\cite{CMS_PAS_TOP_11_009}. Then we can
calculate the invariant mass of the fat jet and the reconstructed
$W$ boson.

\begin{widetext}

\begin{figure}[h]                                 %Fig.12
\centering   \bc
\includegraphics[width=10cm,height=10cm]{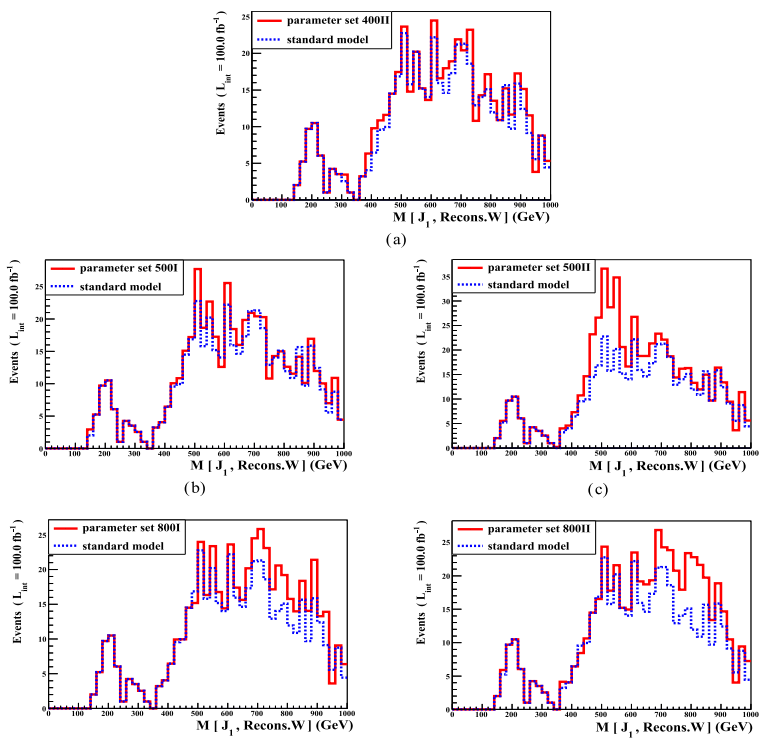}
 \ec
 \vspace{-1cm} \caption{Invariant mass $M(J_1,recons.W)$ distributions (red-solid) for the five examples [(a) 400II, (b) 500I, (c) 500II, (d) 800I,
 (e) 800II]
 together with that of the SM (blue-dotted) for comparison,
 with ${\sf L}^{}_{int}=$100 fb$^{-1}$.}
 \label{invariant-mass-distribution}                       %
\end{figure}
\end{widetext}

In Fig.\,\ref{invariant-mass-distribution}, we plot the invariant
mass $M(J_1,recons.W)$ distributions (red-solid) for five
examples, together with that of the SM distribution (blue-dotted)
for comparison, with an integrated luminosity of 100 fb$^{-1}$.

 We
see that there are excess events over the SM results around
$M^{}_H$. This can be a signal of the contribution of the
intermediate state heavy Higgs boson. So observation of the excess
events can be a way of discovering the heavy Higgs boson.
Comparing the five distributions in
FIG.\,\ref{invariant-mass-distribution}, we see that the excess
events are more significant for heavier $H$ than for lighter $H$.

\newpage

\section{Probing Heavy Neutral Higgs Bosons via $\bm{pp\to VH^\ast\to VVV}$
associated production}

Now we study the process $pp\to VH^\ast\to VVV\to
\ell^+\nu^{}_\ell j^{}_1j^{}_2j^{}_3j^{}_4$, $V=W, Z$. Here the
$W$ boson decaying to $\ell^+\nu^{}_\ell$ can be either the weak
boson associated with $H$ or a weak boson in $H$ decay. The other
two weak bosons decay to $j^{}_1j^{}_2j^{}_3j^{}_4\sim
J^{}_1J^{}_2$, where $J^{}_1$ and $J^{}_2$ are two fat jets. From
now on, we take a convention regarding $J^{}_1$ as the fat jet
with largest transverse momentum, and $J^{}_2$ as the one with
second largest transverse momentum.

\begin{widetext}

\begin{figure}[h]                                 %Fig.13
\centering   \bc
\includegraphics[width=16cm,height=8.5cm]{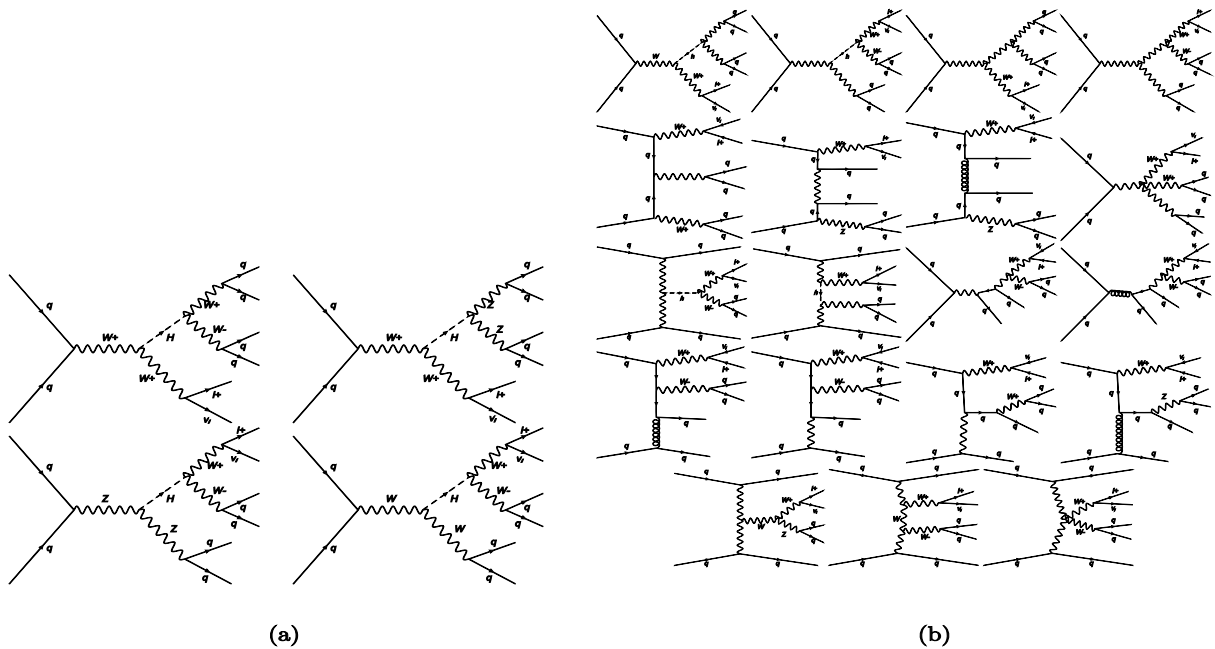}
 \ec \vspace{-0.4cm} \caption{Feynman diagrams in $VH$ associated production. (a) The
 signal, (b) examples of the IB.}
 \label{VH-S+IB}                       %
\end{figure}
\end{widetext}

The Feynman diagrams for the signal and example of the IB are
shown in FIG.\,\ref{VH-S+IB}
Again, these two amplitudes have interference, so that they should
be calculated together.

Next we consider the RBs. Now the largest QCD background is the
inclusive $W+2j$ when $W\to \ell^+\nu_\ell$ and the two jets mimic
the two fat jets in the signal. For safety, we take into account
all the $W+$ jets and the $W+V+$ jets processes to do the
simulation, and pick up the parts that can mimic the signal as the
QCD backgrounds. For the top-quark background, we make the same
treatment (cf. FIG.\,\ref{topB}).

We then make the following kinematic cuts for suppressing the
backgrounds.

\null\noindent {\bf cut1}: {\it Leptonic cuts}

Similar to what we did in Sec.\,II, we require an isolated
$\ell^+$ ($\mu^+,e^+$) in the detectable rapidity region (cf.
TABLE\,\ref{acceptance}), i.e.,
\bea                                 %((49)
&&N(\ell^+)=1,~~~~N(\ell^-)=0\nonumber\\
{\rm with}~~~~~~~~~~~~&&\eta^{}_{\ell^+}<2.4.
\label{VH-eta_l}                  %
\eea                                %
\begin{figure}[h]                                 %Fig.14
\centering   \bc
\includegraphics[width=5.5cm,height=3cm]{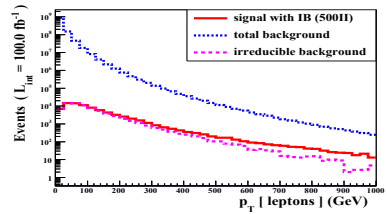}\\
 \ec \vspace{-0.4cm} \caption{$p^{}_T(leptons)$ distributions of signal+IB (red-solid),
 IB (pink-dotted) and total RBs (blue-small-dotted) in $pp\to VH^\ast\to VVV$ process for example 500II with ${\sf L}^{}_{int}=$100 fb$^{-1}$.}
 \label{leptonp_T}                       %
\end{figure}

Next we make the cut on $p^{}_T(leptons)$. The inclusive
$p^{}_T(leptons)$ distributions of the signal plus IB, the RB and
the total background are shown in FIG.\,\ref{leptonp_T}. Here, we
do not have to take care of the transverse momentum balance with
the forward jets as in Sec.\,II, so we can take a stronger cut
\bea                           %(50)
\left|p^{}_T(leptons)\right|>400\,{\rm GeV}
\label{VH-leptonp_T}                    %
\eea                                     %
to suppress more backgrounds. This cut can also strongly suppress the fake leptons.\\

\null\noindent {\bf cut2}: {\it fat jet cuts}

As mentioned in Sec.\,II, we require the first two large
transverse momenta to satisfy
\bea                                 %(51)
&&70\,{\rm GeV}<M(J^{}_1)<100\,{\rm GeV}\nonumber\\
&&70\,{\rm GeV}<M(J^{}_2)<100\,{\rm GeV}.
\label{J_1J_2}                                %
\eea                                           %
This can
suppress the backgrounds with ordinary jets.\\

\null\noindent {\bf cut3}: {\it Top-quark veto}

As in Sec.\,II, for suppressing the top-quark background, we
construct two invariant masses $M(J,j^{}_1)$ and $M(J,j^{}_2)$,
where $J=J^{}_1~{\rm or}~J^{}_2$, and $j^{}_1$, $j^{}_2$ are the
two observed jets from the partons $b$ or $\bar b$ in
FIG.\,\ref{topB}. In FIG.\,\ref{VH-M(J,j)} we plot the
$M(J,j^{}_1)$ [or $M(J,j^{}_2)$] distribution from our simulation
including the signal plus IB (red-solid) and the top-quark
background (blue-dotted) distributions for the example 500II with
${\sf L}^{}_{int}=$100 fb$^{-1}$. We can see clearly the top-quark
peak (in the blue-dotted curve) in the region $130\,{\rm
GeV}<M(J,j)<240\,{\rm GeV}$ for $j$=$j^{}_1$ or $j^{}_2$.

\begin{figure}[h]                                 %Fig.15
\centering   \bc
\includegraphics[width=5.5cm,height=3cm]{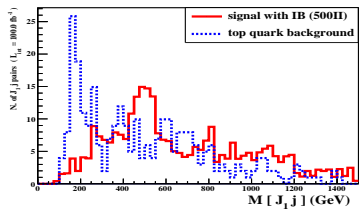}\\
 \ec \vspace{-0.4cm} \caption{$M(J,j)$ distributions of signal+IB (red-solid)
 and the top-quark background (blue-dotted) in $pp\to VH^\ast\to VVV$ process for the example 500II with ${\sf L}^{}_{int}=$100 fb$^{-1}$.}
 \label{VH-M(J,j)}                         %
\end{figure}
As in Sec.\,II, we set the cut
\bea                               %(52)
{\rm 130\,GeV}<M(J,j)<{\rm 240\,GeV},
\label{topveto'}                           %
\eea                                       %
to suppress the top-quark background. We should veto the event if
one of $M(J,j^{}_1)$ and $M(J,j^{}_2)$ satisfies (\ref{topveto'}).
Equivalently, we only take the events in which both $M(J,j^{}_1)$
and $M(J,j^{}_2)$ are outside the region (\ref{topveto'}). In this
way, we can effectively veto the top-quark background events.\\

\null\noindent {\bf cut4}: {\it The $\Delta R(\ell^+,J^{}_1,)$
cut}

\begin{figure}[h]                                 %Fig.16
\centering   \bc
\includegraphics[width=6cm,height=3cm]{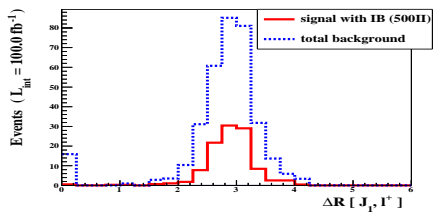}\\
 \ec \vspace{-0.4cm} \caption{$\Delta R(\ell^+,J^{}_1)$ distributions of signal+IB (red-solid)
 and the total background (blue-dotted) in the $pp\to VH^\ast\to VVV$ process for example 500II with ${\sf L}^{}_{int}=$100 fb$^{-1}$.}
 \label{VH-DeltaR}                         %
\end{figure}

In  $VH$ associated production, because $H$ is heavy and has a
quite large momentum, the recoil transverse momentum of the
associated $V$ boson is generally large. Furthermore, due to the
large momentum of the heavy Higgs boson $H$, the angular distance
between two weak bosons from $H$ decay is small, while that
between the weak boson associated with $H$ and any of the weak
boson in $H$ decay is large. If $\ell^+$ comes from the $W$ boson
associated with $H$, the angular distance between $\ell^+$ and any
of the fat jets is large. If $\ell^+$ comes from the decay of $H$,
there must be a fat jet $J_1$ (actually from the $V$ boson
associated with $H$) with large $\Delta R(\ell^+,J^{}_1)$. The
background does not have this situation. We plot, in
FIG.\,\ref{VH-DeltaR}, the $\Delta R(\ell^+,J^{}_1)$ distributions
of the signal plus IB (red-solid) and the total background
(blue-dotted) in the $VH$ associated production for the example
500II with ${\sf L}^{}_{int}=$100 fb$^{-1}$. We see that the main
distribution of the red-solid curve is indeed located further
right to that of the blue-dotted curve. So that a cut

\bea                                   %(53)
\Delta R(\ell^+,J^{}_1)>2.5
\label{DeltaRcut}                           %
\eea                                        %
can suppress the total background.

We know that {\bf cut1} on the leptons can effectively avoid the
fake leptons from ordinary jets to mimic the signal lepton.
However, since the fat jets $J^{}_1$ and $J^{}_2$ have quite large
transverse momenta, {\bf cut1} may not be sufficient to suppress
the fake leptons from the fat jets. Therefore, we should require
the lepton not to overlap with any of the fat jets. Since we have
taken $R=0.7$ in jet formations, this means both $\Delta
R(\ell^+,J^{}_1)$ and $\Delta R(\ell^+,J^{}_2)$ should be larger
than 0.7. {\bf cut4} already guarantees $\Delta R(\ell^+,J^{}_1)$
to satisfy this requirement. So that we add the requirement
\bea                           %(54)
\Delta R(\ell^+,J^{}_2)>0.7
\label{DeltaR(l,J_2}               %
\eea                              %
here.\\

\begin{widetext}

\begin{table}[h]                  %Tab5
 \tabcolsep 10pt
\begin{center}
\caption{Cut efficiencies expressed in terms of the tree-level
cross sections $\sigma^{}_{S+IB}$ and $\sigma^{}_B$ (in units of
fb) in the $pp\to VH^\ast\to VVV$ process. The first five columns
are values of $\sigma^{}_{S+IB}$ for the five examples, and The
last four columns are values $\sigma^{}_B$ for four kinds of
backgrounds.}
\begin{tabular}{ccccccccccc}
\hline \hline
& & &$\sigma^{}_{S+IB}$& & & & &&$\sigma^{}_B$& \\
& 400\uppercase\expandafter{\romannumeral2}  &
500\uppercase\expandafter{\romannumeral1} &
500\uppercase\expandafter{\romannumeral2}
& 800\uppercase\expandafter{\romannumeral1} & 800\uppercase\expandafter{\romannumeral2} && IB &  W+jets & $t\bar{t}$ & WV+jets \\
\hline
without cuts &2085  &2037  &2009  &1917  &1996  &&1925  &31500000&92000 &7600  \\
Cut 1   &46.9  &54.4  &25.7  &18.6  &25.3  &&13.1  &1422    &65.9  &47.9  \\
Cut 2   &2.78  &4.36  &1.21  &0.629 &1.41  &&0.211 &2.91    &0.716 &0.336 \\
Cut 3   &2.32  &3.79  &1.08  &0.526 &1.24  &&0.13  &2.15    &0.149 &0.25  \\
Cut 4   &2.04  &3.21  &0.921 &0.426 &1.11  &&0.061 &1.39    &0.060 &0.179 \\
\hline \hline
\end{tabular}
\label{VH-efficiency}
\end{center}
\end{table}
\end{widetext}

\begin{center}
\begin{table}[h]                            %Tab6
%\label{VV-IL}                    %
\caption{Required integrated luminosity ${\sf L}^{}_{int}$ (in
units of fb$^{-1}$) for the statistical significance of
$1\sigma,\,3\sigma$ and $5\sigma$ for the five examples in $pp\to
VH^\ast\to VVV$ process.}
\tabcolsep 9pt                        %
\begin{tabular}{cccccc}
\hline \hline
 & & &${\sf L}_{int}$\,[fb$^{-1}_{}]$ & &\\
 &400II&500I&500II&800I&800II\\
 \hline
$1\sigma$&0.43 &0.18 &2.3 &13 &1.6 \\
$3\sigma$&3.9 &1.6 &21 &115 &14 \\
$5\sigma$&10.8 &4.5 &57 &319 &39 \\
 \hline \hline
\end{tabular}
\label{VH-IL}
\end{table}
\end{center}

To see the efficiency of each cut, we list the values of the cross
sections (in fb) for signal plus IB (for the five examples
mentioned in Sec.\,I) and various backgrounds after each cut in
TABLE\,\ref{VH-efficiency}. We see that, with all these cuts, the
backgrounds can be effectively suppressed. Compared with the
numbers in TABLE\,\ref{VV-efficiency}, we see that all the
backgrounds in TABLE\,\ref{VH-efficiency} are more suppressed.
Again the signal plus IB cross section is of the order of 0.4--3
fb, so that for an integrated luminosity of around 100 fb$^{-1}$,
we can have a few tens to a few hundreds of events.

From Eqs.\,(\ref{sigma})--(\ref{stat}), we obtain the required
integrated luminosity for the statistical significance of
$1\sigma$, $3\sigma$ and $5\sigma$ for the five examples mentioned
in Sec.\,I (cf. TABLE\,\ref{VH-IL}).

We see that, except for 800I, all the other four examples are
hopeful to be discovered ($5\sigma$) in the first few years run of
the 14 TeV LHC; while 800I can have evidence (3$\sigma$) for ${\sf
L}_{int}=115$ fb$^{-1}$, and can be discovered ($5\sigma$) for
${\sf L}_{int}=319$ fb$^{-1}$ at the 14 TeV LHC. These are
conclusions considering only the statistical errors.
\begin{widetext}

\begin{figure}[h]                                 %Fig.17
\centering \bc
\includegraphics[width=12cm,height=9cm]{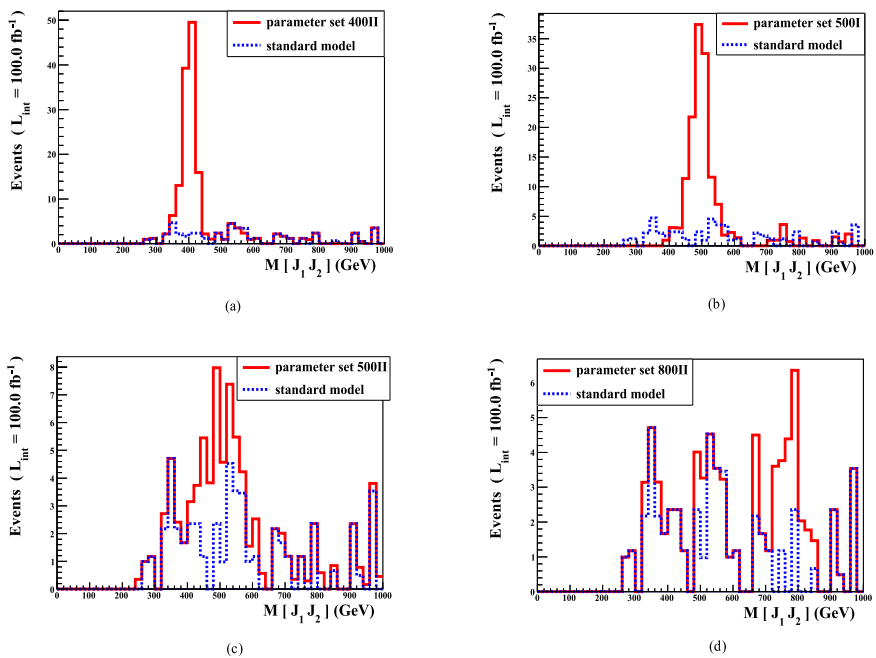} \ec
 \vspace{-0.6cm}
 \caption{Invariant masses $M(J^{}_1,J^{}_2)$ in the $pp\to VH^\ast\to VVV$ process after all cuts and $\Delta R(\ell^+,J^{}_2)>2.5$ for the examples:
 (a) 400II, (b) 500I, (c) 500II, and (d) 800II.}
\label{VH-M}                       %
\end{figure}
\end{widetext}

Finally, we deal with the issue of experimentally discovering $H$
and measuring $M^{}_H$. In addition to {\bf cut4}, we add a cut
\bea                                 %(55)
\Delta R(\ell^+,J^{}_2)>2.5,             %
\label{R(l,J_2)}             %
\eea                          %
where $J^{}_2$ is the other fat jet. Then both $J^{}_1$ and
$J^{}_2$ will mainly come from the decay of $H$, and thus the
invariant mass $M(J^{}_1,J^{}_2)$ will show the $H$ peak at
$M(J_1,J_2)=M^{}_H$. Since the uncertainties in identifying the
fat jet from a boosted W boson decay are small
\cite{CMS_JME_13_006}, measuring the $M(J^{}_1,J^{}_2)$
distribution is quite feasible experimentally.

 FIG.\,\ref{VH-M} shows the
$M(J^{}_1,J^{}_2)$ distributions for examples 400II, 500I, 500II
and 800II. We see that sharp peaks can be seen clearly, and thus
the heavy Higgs boson and its mass can be detected experimentally.
This is the advantage of the $pp\to VH^\ast\to VVV$ process.

\begin{figure}[h]                                 %Fig.18
\centering   \bc
\includegraphics[width=5cm,height=3.5cm]{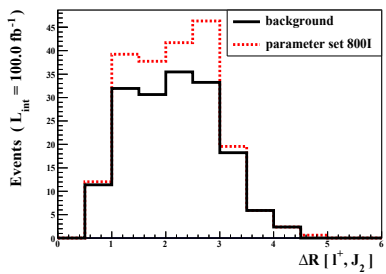}\\
 \ec \vspace{-0.4cm}             %
 \caption{$\Delta R(\ell^+,J^{}_2)$ distributions of signal+IB (red-dotted)
 and the total background (dark-solid) in the $pp\to VH^\ast\to VVV$ process for the example 800I with ${\sf L}^{}_{int}=$100 fb$^{-1}$.}
 \label{VH-DeltaRlJ800I}                         %
\end{figure}

The example 800I is special. It has a very large decay width due
to the largeness of $\Gamma(H\to t{\bar t})$, so that there cannot
be a sharp peak showing up. However, due to the fact that
$M^{}_H\gg M^{}_h$ in this example, the heavy Higgs boson $H$
moves much more slowly than the light Higgs boson $h$ does.
Therefore, $\Delta R(\ell^+,J^{}_2)$ for $H$ is larger than that
for $h$ in the SM background. In FIG.\,\ref{VH-DeltaRlJ800I} we
plot the $\Delta R(\ell^+,J^{}_2)$ distributions of the signal
plus IB (red-dotted) and the SM background (dark-solid) in the
range $\Delta R(\ell^+,J^{}_2)>0.7$ due to (\ref{DeltaR(l,J_2}).

We see that the main distribution of the signal plus IB is located
around $\Delta R(\ell^+,J^{}_2)=2.7$ which is right to that of the
SM background at around $\Delta R(\ell^+,J^{}_2)=2.3$, and the
height of signal plus IB is higher. This can be seen as a
characteristic feature of the heavy Higgs boson contribution in
example 800I.

In principle, we can replace the cut (\ref{R(l,J_2)}) by $\Delta
R(J_1,J_2)>2.5$ to extract the contribution of the Feynman diagram
in which the leptons are from $H$ decay, and use the
reconstruction method suggested in Ref.\,\cite{Butterworth} to
calculate the invariant mass $M(J_1,recons.W)$ distribution as
what we did in FIG.\,\ref{invariant-mass-distribution}. However,
our result shows that the obtained resonance peaks are less clear
than those in FIG.\,\ref{VH-M}. So we only suggest the method
presented above.

From FIG.\,\ref{VH-M} we see that the excess of events over the SM
result is more significant for lighter Higgs boson than for
heavier Higgs boson. This is just the opposite to that in the $VV$
scattering process (cf. the last paragraph in Sec.\,II). This
means that {\it the $VV$ scattering process and the $pp\to
VH^\ast\to VVV$ process are complementary to each other in this
respect.}

Having found the resonance, the next task is to determine whether
its spin is really zero. This can be done by studying the decay
mode $H\to ZZ\to 4\ell$ \cite{Miller} which needs much larger
integrated luminosity. Another possible way is to measure the
azimuthal angle dependence as suggested by
ref.\,\cite{Murayama}.\\\\

\section{Measuring the Anomalous Coupling Constants $\bm{f^{}_W}$
and $\bm{f^{}_{WW}}$}

If we can measure the values of the anomalous coupling constants
$f^{}_W$ and $f^{}_{WW}$ which characterize the heavy neutral
Higgs boson $H$, it will be a new high energy measurement of the
property of the nature, and will serve as a new high energy
criterion for the correct new physics model. All new physics
models predicting $f^{}_W$ and $f^{}_{WW}$ not consistent with the
measured values should be ruled out. {\it The necessary condition
for surviving new physics models is that their predicted $f^{}_W$
and $f^{}_{WW}$ should be consistent with the measured values}. We
shall see that this measurement is really possible.

It has been pointed out in Ref.\,\cite{QKLZ09} that, for a
single-Higgs system, measuring both the cross section and the
leptonic transverse momentum distribution in weak-boson scattering
processes may determine the values of $f^{}_W$ and $f^{}_{WW}$ to
a certain precision. However, in our present case with both $h$
and $H$ contributions, the weak-boson scattering process is not so
optimistic for this purpose. So we concentrate on studying the
measurement of $f^{}_W$ and $f^{}_{WW}$ in the $pp\to VH^\ast\to
VVV$ process.\\

\subsection{The Case of $\bm{M^{}_H=500}$ GeV as an Example}

Let us take the case of $M^{}_H=500$ GeV as an example. After
measuring the resonance peak experimentally, we can impose an
additional cut
\bea                               %(56)
400\,{\rm GeV}<M(J_1,J_2)<600\,{\rm GeV}
\label{newM(J_1,J_2)}                           %
\eea                                            %
to take the events in the vicinity of the resonance peak to
further improve the signal to background ratio. Now we take four
sets of the anomalous coupling constants $f^{}_W$ and $f^{}_{WW}$,
and see if there can be certain new observables to distinguish
them. We take\\

\null\noindent {\bf set I}: $C^{}_t=1,\,\rho^{}_h=0.8$,
$\rho^{}_H=0$, and $f^{}_W=f^{}_{WW}=0$\\ \null~~~~ (background).\\
\null\noindent {\bf set II}: $C_t=0.6,\,\rho^{}_h=0.8$,
$\rho_H=0.6$ and\\ \null~~~~ $f^{}_W=-f^{}_{WW}=6$ TeV$^{-2}$.\\
\null\noindent {\bf set III}: $C_t=0.6,\,\rho^{}_h=0.8$,
$\rho_H=0.6$, and $f^{}_W=12$\\ \null~~~~ TeV$^{-2}$$\gg f^{}_{WW}=0$.\\
\null\noindent{\bf set IV}: $C_t=0.6,\,\rho^{}_h=0.8$,
$\rho_H=0.6$, and\\ \null~~~~ $f^{}_W=0\ll f^{}_{WW}=12$
TeV$^{-2}$.\\

 We can now construct several observables
which may be able to distinguish the four sets of $f^{}_W$ and
$f^{}_{WW}$ listed above, namely (a) the $p^{}_T(leptons)$
distribution, (b) the $p^{}_T(J^{}_1)$ distribution, (c) the
$\Delta R(\ell^+,J^{}_2)$ distribution, and (d) the $\Delta
R(J^{}_1,J^{}_2)$ distribution. In the two transverse momentum
distributions, the additional cuts (\ref{newM(J_1,J_2)}) and
$\Delta R(\ell^+,J^{}_2)>2.5$ are taken, while in the two angular
distance distributions none of these additional cuts is taken.

\begin{widetext}

\begin{figure}[h]                                 %Fig.19
\centering   \bc
\includegraphics[width=11cm,height=9cm]{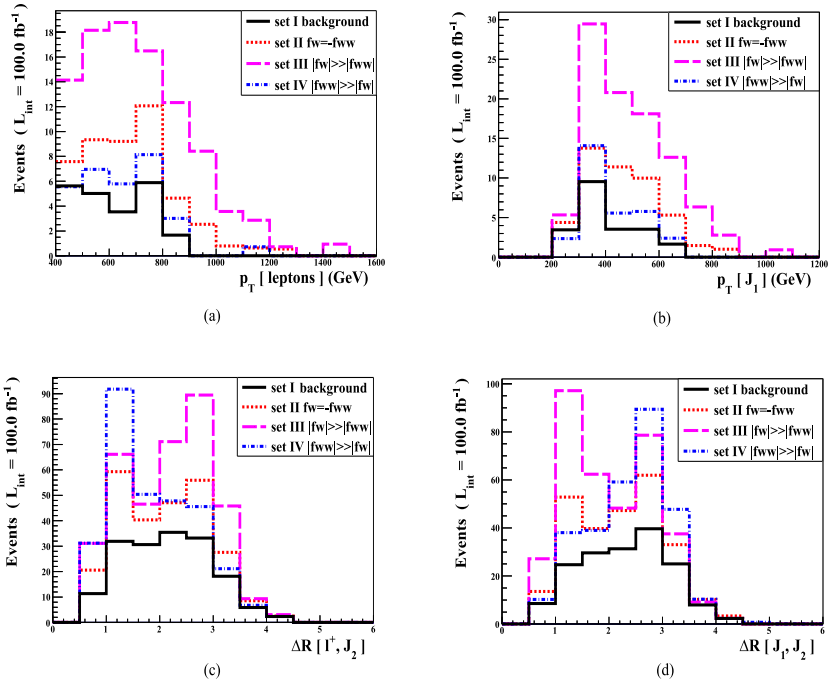}
 \ec \vspace{-0.4cm}             %
 \caption{(a) The $p^{}_T(leptons)$ distribution [with (\ref{newM(J_1,J_2)}) and $\Delta R(\ell^+,J^{}_2)>2.5$] ,
 (b) the $p^{}_T(J^{}_1)$ distribution [with (\ref{newM(J_1,J_2)}) and $\Delta R(\ell^+,J^{}_2)>2.5$], (c) the $\Delta R(\ell^+,J^{}_2)$
distribution [without (\ref{newM(J_1,J_2)}) and $\Delta
R(\ell^+,J^{}_2)>2.5$], and (d) the $\Delta R(J^{}_1,J^{}_2)$
distribution [without (\ref{newM(J_1,J_2)}) and $\Delta
R(\ell^+,J^{}_2)>2.5$], with ${\sf L}_{int}=100$ fb$^{-1}$. The
dark-solid, red-dotted, pink-dashed, and blue-dashed-dotted curves
stand for {\bf set I, set II, set III} and {\bf set IV},
respectively.}
 \label{VH-distributions}                         %
\end{figure}
\end{widetext}

In FIG.\,\ref{VH-distributions} we plot these four distributions
for the four sets of $f^{}_W$ and $f^{}_{WW}$ with ${\sf
L}_{int}=100$ fb$^{-1}$, where the dark-solid, red-dotted,
pink-dashed, and blue-dashed-dotted curves stand for {\bf set I,
set II, set III} and {\bf set IV}, respectively.

We see that, in all the four distributions, the curves of the four
sets can be clearly distinguished. The differences between
different sets in FIG.\,\ref{VH-distributions}(c) and
FIG.\,\ref{VH-distributions}(d) are more significant. Therefore,
{\it measuring the four distributions experimentally, and checking
with each other, the relative size of $f^{}_W$ and $f^{}_{WW}$
existing in the nature can be obtained, and together with the
measurement of the cross section, the values of $f^{}_W$ and
$f^{}_{WW}$ can be separately determined, which gives the new
criterion for discriminating new physics models}. This is an
important advantage of the $pp\to VH^\ast\to VVV$ process.

\subsection{The Case of $\bm{M^{}_H=800}$ GeV as an Example}

\begin{widetext}

\begin{figure}[h]                                 %Fig.20
\centering   \bc
\includegraphics[width=12cm,height=5cm]{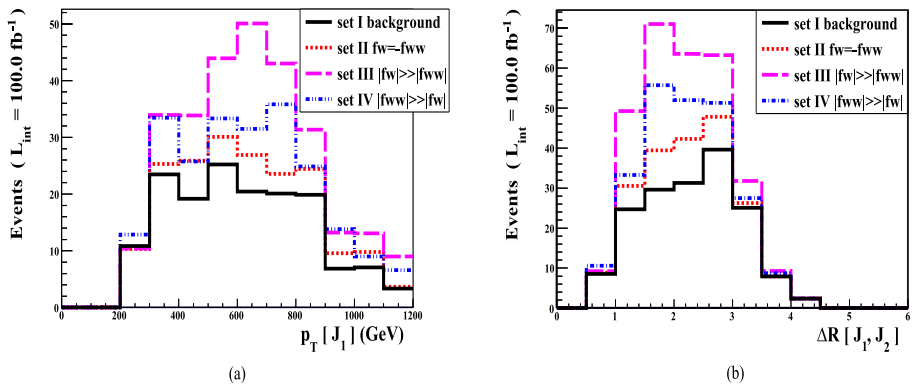}
 \ec \vspace{-0.6cm}                %
\caption{(a) the $p^{}_T(J^{}_1)$ distribution
 and (b) the $\Delta
R(J^{}_1,J^{}_2)$ distribution for $M^{}_H=800$ GeV with ${\sf
L}_{int}=100$ fb$^{-1}$. The meaning of the curves is the same as
in FIG.\,\ref{VH-distributions} but with $C_t=1$.}
 \label{VH-800-distributions}                         %
\end{figure}
\end{widetext}

Since in the case of 800I no clear peak can be seen and it can
only be realized by the distribution in
FIG.\,\ref{VH-DeltaRlJ800I}, we now examine whether it is possible
to measure the values of $f^{}_W$ and $f^{}_{WW}$ in this case. In
FIG.\,\ref{VH-800-distributions} we plot the $p^{}_T(J^{}_1)$ and
$\Delta R(J^{}_1,J^{}_2)$ distributions for the $M^{}_H=800$ GeV
case with four sets of parameters as those in the case of
$M^{}_H=500$ GeV but with $C_t=1$. We see that the four sets of
$f^{}_W$ and $f^{}_{WW}$ can all be clearly distinguished.\\

\section{Summary and Discussion}

To search for new physics beyond the SM, we suggest searching for
heavy neutral Higgs bosons which are generally contained in new
physics models.

We summarize our results as follows.

\begin{description}
\item{i}~~~~ In this paper, we have considered an arbitrary new
physics theory containing more than one Higgs bosons
$\Phi^{}_1,\,\Phi^{}_2,\cdots$ taking account of their mixing
effect. For generality, we do not specify the EW gauge group
except requiring that it contains an $SU(2)_L\times U(1)$ subgroup
with the gauge bosons $W,\,Z$ and $\gamma$. We also neither
specify the number of $\Phi^{}_1,\,\Phi^{}_2,\cdots$, nor specify
how they mix to form mass eigenstates except to identify the
lightest Higgs boson $h$ to the recently discovered
$M^{}_h=$125--126 GeV Higgs boson. Then we study the general
properties of the couplings of both the lightest Higgs boson $h$
and a heavier neutral Higgs boson $H$ (lighter than other heavy
Higgs bosons). The probe of gauge-phobic heavy neutral Higgs
bosons are not considered in this study, and
will be studied elsewhere.\\
\null~~~~ We first gave a general model-independent formulation of
the couplings of $h$ and $H$ to fermions and gauge bosons based on
the idea of the effective Lagrangian up to dim-6 operators in
Sec.\,II. The obtained effective couplings for the Higgs-gauge
interaction are different from the traditional ones constructed
for a single-Higgs system by containing new parameters $\rho^{}_h$
and $\rho^{}_H$ reflecting the Higgs mixing effect. After taking
account of the constraints from the known low energy experiments,
there are seven unknown coupling constants left, namely the gauge
coupling constant $\rho^{}_h$ in the dim-4 gauge interaction of
$h$ [cf. Eq.(\ref{hcoupling})], the gauge coupling constant
$\rho^{}_H$ in the dim-4 gauge interaction of $H$ [cf.
Eq.\,(\ref{dim-4Hcoupling})], the anomalous coupling constants
$f^{}_W,\,f^{}_{WW},\,f^{}_B,\,f^{}_{BB}$ in the dim-6 gauge
interactions of $H$ [cf. (\ref{LHeff}), and (\ref{g})], and the
anomalous Yukawa coupling constant $C_t$ of $H$ [cf.
Eq.\.(\ref{C_f})], and the corresponding momentum representations
are given in Eqs.\,(\ref{Hgammagamma}), (\ref{csHZgamma}),
(\ref{HZgamma}), (\ref{HWW}), (\ref{csHZZ}), and (\ref{HZZ}).

\item{ii}~~~~ To estimate the possible range of the anomalous
coupling constants $f^{}_W,\,f^{}_{WW},\,f^{}_B,\,f^{}_{BB}$, we
first studied the theoretical constrains from the requirement of
the unitarity of the $S$-matrix of weak-boson scattering in
Sec.\,III. We took the effective $W$ approximation to calculate
the scattering amplitudes, and calculate the constraints on
$f^{}_W$ and $f^{}_{WW}$ by a two-parameter numerical analysis.
The obtained constraints are shown in FIG.\,\ref{UB}.

\item{iii}~~~~ We further studied the experimental constraints
from the ATLAS and CMS experiments in Sec.\,IV to obtain further
constraints. Anomalous coupling constants consistent with both the
unitarity constraints and the experimental constraints are the
available anomalous couplings that an existing heavy neutral Higgs
boson can have.\\
 \null~~We first make an approximation of
neglecting the anomalous coupling constants in the $H\gamma\gamma$
and $HZ\gamma$ couplings inspired by the trend of the ATLAS and
CMS measurements of $\mu=\sigma/\sigma_{SM}|^{}_{95\%~CL}$ in the
decay channels $H\to\gamma\gamma$ and $H\to Z\gamma$. This
approximation leads to the constraints (\ref{f_BB-f_WW}) and
(\ref{f_B-f_Wf_WW}) which simplifies our analysis.\\
\null~~Then we consider the CMS exclusion bounds on the SM Higgs
boson for the Higgs mass up to 1 TeV, to obtain the experimental
bounds on $f^{}_W$ and $f^{}_{WW}$. The calculation is to full
leading order in perturbation. We took the cases of 400II, 500I,
and 500II as examples. In our calculation of the total decay width
of $H$, we have made a conservative approximation. The obtained
conservative experimental constraints and the {\it available}
regions of $f^{}_W$ and $f^{}_{WW}$ are shown as the blue shaded
regions in FIGs.\,\ref{EB400II}, \ref{EB500I}, and \ref{EB500II}.
This guarantees that a heavy Higgs boson $H$, with its $f^{}_W$
and $f^{}_{WW}$ in the blue shaded regions, is definitely not
excluded by the CMS exclusion bound \cite{CMS_HIG_13_002}. In the
cases of 800I and 800II, there is almost no experimental
constraint on $f^{}_W$ and $f^{}_{WW}$ because the CMS exclusion
bound is very loose at $M^{}_H=800$ GeV.

\item{iv}~~~~ In this paper, for studying the LHC signatures of
$H$, we suggest taking $VV$ scattering and $pp\to VH^\ast\to VVV$
as sensitive processes for probing the anomalous heavy Higgs boson
model-independently at the 14 TeV LHC. We take the general
model-independent formulation of the heavy Higgs couplings in
Sec.\,II. and take five sets of anomalous coupling constants
allowed by the unitarity constraint and the present CMS
experimental exclusion bound as examples to do numerical
simulation, namely 400II, 500I, 500II, 800I, 800II with the heavy
Higgs mass $M^{}_H=$ 400\,GeV, 500\,GeV, and 800\,GeV (cf.
Sec.\,IV). The calculations are to the hadron level. We take the
CTEQ6.1 parton distribution functions \cite{CTEQ6.1}, and use
MADGRAPH5 \cite{MADGRAPH5} to do the full tree-level simulation.
The parton shower and hadronization are calculated with PYTHIA6.4
\cite{PYTHIA6.4}, and the anti-$k^{}_T$ algorithm with $R=0.7$
\cite{antik_T} in DELPHES 3 \cite{DELPHES3} is used for the
formation of jets. We also use DELPHES 3 to simulate the detecting
efficiency of the detector.

\item{v}~~~~ We first study the the semileptonic decay mode of
weak-boson scattering, i.e., $pp\to VVj^f_1j^f_2\to \ell^+
\nu^{}_\ell j_1j_2j^f_1j^f_2$. The Feynman diagrams of the signal
and backgrounds are shown in FIGs.\,\ref{VV-sig&IB}--\ref{topB}.
The largest background is the QCD background, the inclusive $pp\to
W+3j$ which is larger than the signal plus irreducible background
(IB) by four orders of magnitude. To suppress the backgrounds, we
imposed five kinematic cuts given in
Eqs.\,(\ref{leptoncut})--(\ref{topveto}) which can effectively
suppress the backgrounds. The cut efficiencies of each cut are
listed in TABLE\,\ref{VV-efficiency}, and the required integrated
luminosities for $1\sigma$ deviation, $3\sigma$ evidence, and
$5\sigma$ discovery are shown in TABLE\,\ref{VV-IL}. It shows that
examples 500II and 800II are hopeful to be discovered (at the
$5\sigma$ level) in the first few years run of the 14 TeV LHC.
while 800I can be discovered (at the $5\sigma$ level), and 400I
and 500I can have evidence (at the $3\sigma$ level) for an
integrated luminosity of 300 fb $^{-1}$ at the 14 TeV LHC. We then
took the method of determining the longitudinal momentum of the
neutrino by requiring to reconstruct the $W$ boson mass correctly
\cite{Butterworth}, and with which we calculated the invariant
mass $M(J_1,recons.W)$ distributions as shown in
FIG.\,\ref{invariant-mass-distribution}. We see that there are
evident excesses of events over the SM result around
$M(J_1,recons.W)=M^{}_H$. This can be the signal of the
contribution of the intermediate state heavy Higgs boson. We also
see that the excess of events are more significant for heavier
Higgs boson than for lighter Higgs boson.

\item{vi}~~~~ We then study the semileptonic mode of the $pp\to
VH^\ast\to VVV$ process, $pp\to VH^\ast\to VVV\to
\ell^+\nu^{}_\ell j^{}_1j^{}_2j^{}_3j^{}_4\to \ell^+\nu^{}_\ell
J^{}_1J^{}_2$ ($J^{}_1$ and $J^{}_2$ stand for the fat jets with
largest and second largest transverse momenta, respectively). The
Feynman diagrams for the signal and IB are shown in
FIG.\,\ref{VH-S+IB}. Reducible backgrounds include $W+2$-$jet$,
and the top-quark background similar to those in the weak-boson
scattering process. We also imposed five kinematic cuts in
Eqs.\,(\ref{VH-eta_l})--(\ref{DeltaR(l,J_2}). The cut efficiencies
after each cut are listed in TABLE\,\ref{VH-efficiency} which
shows that all backgrounds are more effectively suppressed. The
required integrated luminosities for $1\sigma$ deviation,
$3\sigma$ evidence, and $5\sigma$ discovery are shown in
TABLE\,\ref{VH-IL}. Except for the example 800I, all the other
four examples are hopeful to be discovered ($5\sigma$ level) in
the first few years run of the 14 TeV LHC; while 800I can have an
evidence (3$\sigma$) for ${\sf L}_{int}=115$ fb$^{-1}$, and can be
discovered ($5\sigma$) for ${\sf L}_{int}=319$ fb$^{-1}$ at the 14
TeV LHC.~In FIG.\,\ref{VH-M}, we plot the invariant mass
distributions $M(J^{}_1,J^{}_2)$ for examples 400II, 500I, 500I,
and 800II, which shows that the resonance peaks for all these four
examples are clearly seen. This makes it possible for the
experimental search for the heavy Higgs boson $H$ and the
measurement of its mass $M^{}_H$. For the example 800I, due to the
large decay rate of $\Gamma(H\to t{\bar t})$, the total decay
width of $H$ is very large such that there is no clear peak
showing up. However, FIG.\,\ref{VH-DeltaRlJ800I} shows a
characteristic feature of the $M^{}_H=800$ GeV Higgs boson in the
$\Delta R(\ell^+,J^{}_2)$ distribution, which can help the
experiment to find out the contribution of the heavy Higgs boson
$H$. We see that the excess of events are more significant for
lighter Higgs boson than for heavier Higgs boson. This is just the
opposite to the case of the $VV$ scattering. So, in this sense,
the $VV$ scattering process and the  $pp\to VH^\ast\to VVV$
process are complementary to each other.\\
\null~~After determining the spin of the resonance, one can
confirm the discovery of a heavy Higgs boson.

\item{vii}~~~~ We also show the possibility of measuring the
values of anomalous coupling constants $f^{}_W$ and $f^{}_{WW}$
experimentally by measuring both the cross section and the
$p^{}_T(leptons)$ distribution, the $p^{}_T(J^{}_1)$ distribution,
the $\Delta R(\ell^+,J^{}_2)$ distribution, and the $\Delta
R(J^{}_1,J^{}_2)$ distribution (cf. FIGs.\,\ref{VH-distributions}
and \ref{VH-800-distributions}). This will be a new measurement of
the property of the nature at high energies, and will serve as a
new high energy criterion for the correct new physics model. All
new physics models predicting $f^{}_W$ and $f^{}_{WW}$ not
consistent with the measured values should be ruled out. {\it The
necessary condition for surviving new physics models is that their
predicted $f^{}_W$ and $f^{}_{WW}$ should be consistent with the
measured values}.
\end{description}

In weak-boson scattering, we imposed the forward-jet cut
$p^{}_T(j^f)>35$ GeV to avoid the pile-up events, while we did not
impose that in $VH$ associated production. This is because the
transverse momenta of all the final state particles are large,
e.g., our simulation shows that $p^{}_T(J_2)>100$ GeV,
$p^{}_T(J_1)>200$ GeV [cf. FIG.\,\ref{VH-distributions}(b)], and
$p^{}_T(leptons)>400$ GeV [cf. Eq.\,(\ref{VH-leptonp_T})].

In all our predictions, only the statistical error is considered.
We leave the study of the systematic error related to the details
of the detectors to the experimentalists. Moreover, with the study
of the jet shape, it may further suppress the backgrounds
\cite{BDRS}\cite{Han09}.

In Ref.\,\cite{Englert} the 1-loop level contribution $gg\to VH$
in the SM was studied, and they showed that, although it is
smaller than the tree-level quark initiated contribution, this
contribution can help to enhance the signal in $VH$ associated
production. This may also enhance the signal in our $pp\to
VH^\ast\to VVV$ process. However, in our Type-II examples,
$C_t<1$, so that the gluon initiated contribution is less
important.

\begin{widetext}

\begin{figure}[h]                                 %Fig.21
\centering   \bc
\includegraphics[width=15.2cm,height=4cm]{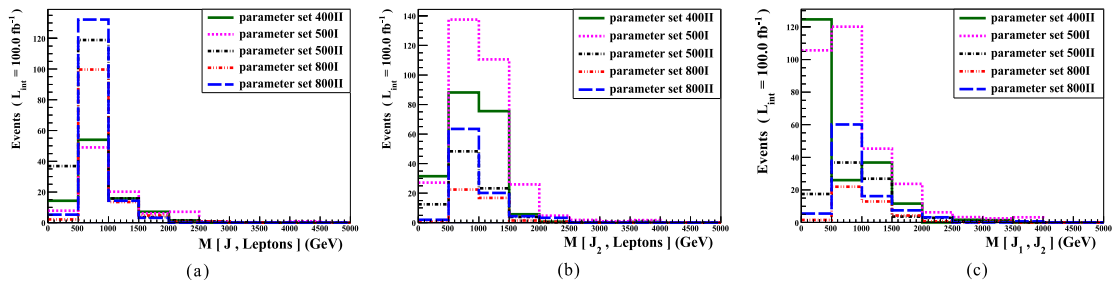}
 \ec \vspace{-0.4cm} \caption{Check of unitarity: (a) $M(J,leptons)$ distribution in weak-boson scattering,
 (b) $M(J^{}_2,leptons)$ distribution and (c) $M(J^{}_1,J^{}_2)$ distribution in the $pp\to VH^\ast\to VVV$
 process.}
 \label{unitarity}                         %
\end{figure}
\end{widetext}

Finally we make a check of the unitarity of our calculation. We
know that the values of the anomalous couplings $f^{}_W$ and
$f^{}_{WW}$ which we take in this paper are consistent with the
unitarity constraints (FIG.\,\ref{UB}). However, the unitarity
constraints are obtained in the effective $W$ approximation. Here
we make a more realistic check based on our full simulation. In
FIG.\,\ref{unitarity}, we plot three invariant mass distributions
up to a few TeV at the LHC in the $VV$ scattering and the $pp\to
VH^\ast\to VVV$ processes. We see that, in the high energy region,
all distributions are monotonically decreasing to zero. This shows
that there is no unitarity violation, so that our calculation is
consistent with the unitarity requirement.
 %, i.e., our calculation
%really makes sense.

\null\noindent {\bf Acknowledgement}  We are grateful to Xin Chen
for valuable discussions. We would also like to thank Tsinghua
National Laboratory for Information Science and Technology for
providing their computing facility. This work is supported by the
National Natural Science Foundation of China under the grant
numbers 11135003 and 11275102.

\null\vspace{1cm}

\bibliography{000}

\begin{thebibliography}{0}
\bibitem{Yukawa}
%The dim-4 Yukawa coupling is momentum-independent which will not
%increase at high energies.
Let us consider if higher dimension
operators can be introduced. We first look at the dim-5 operator
$\bar\psi_f\gamma^\mu D_\mu H\psi_f=\bar\psi_f\gamma^\mu
\partial_\mu H\psi_f+\cdots$. We can write it as
$\partial^{}_\mu(\bar\psi_f\gamma^\mu
H\psi_f)-(\partial_\mu\bar\psi_f)H\psi_f-\bar\psi_f\gamma^\mu\partial_\mu\psi_f$.
The first term (total derivative) contributes only on the surface
at infinity so that can be dropped. For the second and third
terms, the Dirac equation $(\gamma^\mu\partial_\mu-m)\psi_f=0$
reduces them to $m\bar\psi_f H\psi_f$ which is just the Yukawa
form. So the dim-5 operator is actually equivalent to the dim-4
operator for on-shell fermions. It has been argued that the dim-6
operators also do not lead to new forms \cite{A_S}, as we are not
interested in the multi-Hggs-fermion couplings which is irrelevant
to our study. This is why we only take the Yukawa form here.
\end{thebibliography}

\end{document}